\documentclass[aps,prb,twocolumn]{revtex4-2}
\usepackage{tabularx} 
\usepackage{amsmath}  
\usepackage{amssymb}
\usepackage{graphicx} 
\usepackage{hyperref} 
\usepackage{siunitx}
\usepackage{bm}

\begin{document}

\title{Efficient method for calculating magnon-phonon coupling from first principles}
\author{Wuzhang Fang}
\affiliation{Department of Materials Science and Engineering, University of Wisconsin, Madison, WI, 53706, USA}
\author{Jacopo Simoni}
\affiliation{Department of Materials Science and Engineering, University of Wisconsin, Madison, WI, 53706, USA}
\author{Yuan Ping}
\thanks{yping3@wisc.edu}
\affiliation{Department of Materials Science and Engineering, University of Wisconsin, Madison, WI, 53706, USA}
\affiliation{Department of Physics, University of Wisconsin, Madison, WI, 53706, USA}
\affiliation{Department of Chemistry, University of Wisconsin, Madison, WI, 53706, USA}


\begin{abstract}
    Linear magnon-phonon coupling hybridizes magnon and phonon bands at the same energy and momentum, resulting in an anticrossing signature. This hybrid quasiparticle benefits from a long phonon lifetime and efficient magnon transport, showing great potential for spintronics and quantum information science applications. In this paper, we present an efficient and accurate first-principles approach for calculating linear magnon-phonon couplings. 
    We first calculate the magnon spectra from linear spin wave theory with spin Hamiltonian and first-principles exchange constants, which compared well with time-dependent density-functional theory. We then obtain the magnon-phonon coupling from the derivative of off-diagonal exchange constants in real space, calculated from the Hellmann–Feynman forces of the spin-constrained configurations, avoiding the use of cumbersome finite-difference methods.
    Our implementation allows calculating coupling coefficients at an arbitrary wave vector in the Brillouin zone in a single step, through Fourier interpolation of real-space supercell calculations.  We verify our implementation through two-dimensional magnetic systems, monolayer $\mathrm{CrI_3}$, in agreement with experiments, and extend its application to monolayer $\mathrm{CrTe_2}$. We emphasize the role of nonmagnetic atoms in superexchange interactions and magnon-phonon coupling, which have been overlooked previously. We suggest effective tuning of magnon-phonon coupling through strain, doping, and terahertz excitations, for spintronics and quantum magnonics applications.
\end{abstract}

\maketitle

\section{Introduction}\label{sec:intro}
Magnons are quasiparticles describing collective spin excitations in magnetic materials, well recognized as promising information carriers in spintronics due to their lower energy consumption and faster operation speed \cite{chumak_magnon_2015}. Recently, non-classical magnon states have been realized, and initialization and readout protocols have been proposed, paving the way for new continuous-variable quantum information science - quantum magnonics \cite{YUAN20221}. Magnons, when in a bath environment, may interact with other quasiparticles, once energy and momentum requirements have been met.
For example, magnons can couple with phonons due to their similar energy range. The coupling includes both linear magnon-phonon coupling (or magnon-phonon hybridization) and higher-order magnon-phonon interactions, which involve the scattering of magnons with phononic excitations of the lattice. Such scattering is an important mechanism for magnetic damping \cite{Streib2019,Cong2022} and quantum magnonic state dephasing. Magnons and phonons hybridize at the same momentum and energy, resulting in a characteristic anti-crossing feature, which has been observed experimentally in various systems, from single nanomagnets (Ni) \cite{berk_strongly_2019}, to two-dimensional antiferromagnets including $\mathrm{FePS_3}$ \cite{Liu_magnon-phonon_2021,Vaclavkova2021}, $\mathrm{FePSe_3}$ \cite{cui_chirality_2023,luo_evidence_2023} and $\mathrm{MnPSe_3}$ \cite{mai_magnon-phonon_2021}. The gap opening between the anti-crossing bands is a measure of the coupling strength, influencing the rate of energy transfer between magnons and phonons \cite{li_hybrid_2020,hu_design_2024}. The hybrid quasiparticle benefits from a long phonon lifetime and efficient magnon transport, and the coupling is tunable in artificial hybrid magnon-phonon crystals \cite{liao_hybrid_2024}, showing great potentials for spintronic devices \cite{li_hybrid_2020}. 

Linear magnon-phonon coupling in the long-wavelength regime is often described by a phenomenological model with a magnetoelastic coupling constant \cite{Kittel_RMP}. This model can describe the coupling between acoustic magnon and phonon modes, but it cannot describe the coupling between optical magnon and phonon modes and is restricted to the long-wavelength limit. The linear coupling between all the magnon and phonon modes has a matrix form and is a function of the $\bm{q}$ vector in the Brillouin zone (BZ). There have been a few attempts to calculate the linear coupling constant from first principles. In Ref.~\cite{cui_chirality_2023}, the linear coupling in $\mathrm{FePSe_3}$ at $\Gamma$ point is calculated from the difference in the off-diagonal exchange constant between the equilibrium structure and the distorted structures along phonon eigenvectors. In Ref.~\cite{Delugas2023}, the linear coupling constant in $\mathrm{CrI_3}$ at K point is derived from the induced magnetization by a phonon distortion. However, both methods apply to phonon-distorted structures specific to a certain phonon mode, and are restricted to a single $\bm{q}$ vector in the BZ. Our objective is to develop an efficient approach to compute the coupling matrix between all magnon and phonon modes at any arbitrary $\bm{q}$ vector across the BZ, in a single step. 

In this paper, we present a first-principles method to achieve this, and apply it to the monolayer $\mathrm{CrI_3}$, a two-dimensional ferromagnet with a large magnetic anisotropy \cite{huang_layer-dependent_2017}, and monolayer $\mathrm{CrTe_2}$, another two-dimensional ferromagnet with a high Curie temperature \cite{zhang_room-temperature_2021}. We start by calculating the magnon properties using linear spin-wave theory (LSWT), and compare the results with time-dependent density functional theory (TDDFT) using the same exchange-correlation functional. We then derive an expression for the coupling matrix, discuss practical aspects of its implementation, benchmark our implementation with monolayer $\mathrm{CrI_3}$, and extend its application to monolayer $\mathrm{CrTe_2}$.

\section{Theory}
\subsection{Magnon Hamiltonian}\label{sec:mag}
\noindent
We consider a spin Hamiltonian of the following form
\begin{equation}
    H = -\frac{1}{2}\sum_{i\neq j}\sum_{\alpha, \beta}S_i^{\alpha}\cdot \mathcal{J}_{ij}^{\alpha \beta}\cdot S_j^{\beta} -\sum_i K(S_i^z)^2 ,
\end{equation}
where $i,j$ denote the indices of magnetic atoms, $S_i$ is the spin vector, $\mathcal{J}_{ij}^{\alpha\beta}$ is the exchange constant tensor with $\alpha, \beta$ denoting Cartesian directions $x,y,z$, and $K$ is the single-ion magnetic anisotropy constant with a positive value defining an easy axis. We rewrite the Hamiltonian with spin raising and lowering operators $S_i^\pm=S_i^x\pm iS_i^y$ as the sum of two contributions
\begin{equation}
    H = H_0 + H_1
\end{equation}
where
\begin{align}\label{Eq:H0}
    H_0 =& -\frac{1}{2}\sum_{i\neq j}\mathcal{J}_{ij}^{zz} S_i^z S_j^z\nonumber\\
    &-\frac{1}{8}\sum_{i\neq j}\big(\mathcal{J}_{ij}^{+-}S_i^+ S_j^- + \mathcal{J}_{ij}^{-+}S_i^- S_j^+\big)\nonumber \\
    &-\sum_i K(S_i^z)^2,
\end{align}
and
\begin{align}\label{Eq:H1}
    H_1 =& -\frac{1}{4}\sum_{i\neq j}\big(S_i^+\mathcal{J}_{ij}^{-z} + S_i^-\mathcal{J}_{ij}^{+z}\big)S_j^z\nonumber\\ 
    &-\frac{1}{4}\sum_{i\neq j}S_i^z\big(\mathcal{J}_{ij}^{z-}S_j^+ + \mathcal{J}_{ij}^{z+}S_j^-\big).
\end{align}
We have neglected terms of type $S_i^- S_j^-$ and $S_i^+ S_j^+$ and introduced a new set of coefficients to simplify the notation: $\mathcal{J}_{ij}^{+-}=\mathcal{J}_{ij}^{xx}+\mathcal{J}_{ij}^{yy}+i(\mathcal{J}_{ij}^{xy}-\mathcal{J}_{ij}^{yx})$, $\mathcal{J}_{ij}^{-+}=(\mathcal{J}_{ij}^{+-})^*$; $\mathcal{J}_{ij}^{z\pm}=\mathcal{J}_{ij}^{zx}\pm i\mathcal{J}_{ij}^{zy}$, $\mathcal{J}_{ij}^{\pm z}=\mathcal{J}_{ij}^{xz}\pm i\mathcal{J}_{ij}^{yz}$.

The asymmetric part of the off-diagonal exchange constants (in terms of Cartesian directions) gives the components of Dzyaloshinskii–Moriya interaction (DMI) vector, i.e., $D_{ij}^z=(\mathcal{J}_{ij}^{xy}-\mathcal{J}_{ij}^{yx})/2$ \cite{Xiang2013,Toth_2015}. The systems we study in this work are centrosymmetric so that $\mathcal{J}_{ij}^{\alpha\beta}=\mathcal{J}_{ij}^{\beta\alpha}$. As a consequence, the DMI can be set to zero. This simplifies further the expression with $\mathcal{J}_{ij}^{+-}=\mathcal{J}_{ij}^{-+}=\mathcal{J}_{ij}^{xx}+\mathcal{J}_{ij}^{yy}$, and $\mathcal{J}_{ij}^{z\pm}=\mathcal{J}_{ij}^{\pm z}$.
$H_1$ does not contribute to the magnon Hamiltonian since it produces terms $\hat{a}_i$. Furthermore terms beyond second order in the Hamiltonian expansion are neglected (more details are reported in the SI).  

We now perform the Holstein-Primakoff (HP) transformation \cite{HP} on the $H_0$ to obtain the magnon Hamiltonian
\begin{equation}
    H_{\mathrm{mag}} = S\sum_{ij}(\mathcal{J}^{zz}_{ij}+2K)\hat{a}^{\dagger}_i\hat{a}_i - \frac{S}{4}\sum_{ij}\mathcal{J}^{+-}_{ij}(\hat{a}_i \hat{a}^{\dagger}_i+\hat{a}^{\dagger}_i\hat{a}_i).
\end{equation}
The HP transformation is not performed exactly; in practice we linearize $S^+$ and $S^-$, which is known as linear spin wave theory (LSWT). This approximation is justified when the spin wave excitation $\langle\hat{a}^{\dagger}_i\hat{a}_i \rangle$ is much smaller than $2S$, which is the case for the systems considered here.
We further take the $\mathcal{J}_{ij}^{xx}=\mathcal{J}_{ij}^{yy}=\mathcal{J}_{ij}^{zz}=J_{ij}$ and the magnon Hamiltonian becomes
\begin{equation}
    H_{\mathrm{mag}} = S\sum_{ij}(J_{ij}+2K)\hat{a}^{\dagger}_i\hat{a}_i - \frac{S}{2}\sum_{ij}J_{ij}(\hat{a}_i \hat{a}^{\dagger}_i+\hat{a}^{\dagger}_i\hat{a}_i).
\end{equation}
After we perform Fourier transform on the magnon annihilation and creations operators $\hat{a}$ and $\hat{a}^{\dagger}$, we obtain the following magnon Hamiltonian in momentum space
\begin{align}\label{eq:Hmag}
    H_{\mathrm{mag}} &= S\sum_{\bm{k}}\sum_{\bm{L}}\sum_{st}(J_{st}(\bm{L})+2K)\hat{a}^{\dagger}_{s\bm{k}}\hat{a}_{s\bm{k}} \nonumber\\
    &-\frac{S}{2}\sum_{\bm{k}}\sum_{\bm{L}}\sum_{st}J_{st}(\bm{L})e^{-i\bm{k}\cdot\bm{L}}\hat{a}_{s\bm{k}}\hat{a}^{\dagger}_{t\bm{k}} \nonumber\\
    &-\frac{S}{2}\sum_{\bm{k}}\sum_{\bm{L}}\sum_{st}J_{st}(\bm{L})e^{i\bm{k}\cdot\bm{L}}\hat{a}^{\dagger}_{s\bm{k}}\hat{a}_{t\bm{k}}.
\end{align}
Here we change the index $i=(n,s)$ and $j=(m,t)$, where $n,m$ denote the index of unit cell, and $s,t$ denote the index of magnetic atom. We use $J_{st}(\bm{L})$ to represent $J_{ns,mt}$ where $\bm{L}=\bm{R}_m-\bm{R}_n$. Eq. (\ref{eq:Hmag}) can be diagonalized to get the magnon energy and eigenstates. 
 
\subsection{Linear magnon-phonon coupling}\label{sec:coupling}
Linear magnon-phonon coupling is induced by the off-diagonal components of the exchange constant tensor. A proper treatment of such an effect requires a spin noncollinear description and the inclusion of spin-orbit coupling (SOC). The Hamiltonian $H_1$ in Eq.~(\ref{Eq:H1}) can be expressed under the condition of $\mathcal{J}_{ij}^{\alpha\beta}=\mathcal{J}_{ij}^{\beta\alpha}$ as
\begin{align}
    H_1 &= -\frac{1}{2}\sum_{i\neq j}(\mathcal{J}_{ij}^{-z} S_i^+S_j^z + H.c.)
\end{align}
After we apply the HP transformation at the lowest order we get
\begin{align}
    H_1 \approx -\frac{S}{2}\sum_{i\neq j}(\mathcal{J}_{ij}^{-z}S_i^+ + H.c.),
\end{align}
$H.c.$ denotes the complex conjugate. We expand the Hamiltonian to the first order of lattice displacement $\bm{u}_{lv}$ and denote it as linear magnon-phonon coupling
\begin{align}\label{eq:mag-ph}
    H_{\mathrm{mag-ph}} = -\frac{S}{2}\sum_{ns,mt,lv}\bm{u}_{lv}\nabla_{lv} \mathcal{J}_{ns,mt}^{-z}S_{ns}^+ + H.c.,
\end{align}
where we use $l$ to represent the index of unit cell in addition to $m, n$, and $v$ to represent the index of atom in the unit cell in addition to $s,t$. $s,t$ are here restricted to magnetic atom while $v$ is for every atom in the unit cell. We then expand $\bm{u}_{lv}$ on the basis of phonon eigenstates and $S_{ns}^+$ on the basis of magnon eigenstates
\begin{align}
        \bm{u}_{lv} &= \frac{1}{\sqrt{N}}\sum_{\mu\bm{q}}A^{\mu}_v(\bm{q})\bm{e}_v^{\mu}(\bm{q})e^{i\bm{q}\cdot\bm{R}_l}(\hat{b}_{\mu\bm{q}}+\hat{b}^{\dagger}_{\mu-\mathbf{q}}),\\
        S_{ns}^{+}  &= \sqrt{\frac{2S}{N}}\sum_{\nu\bm{k}}f_s^{\nu}(\bm{k})e^{i\bm{k}\cdot\bm{R}_n}\hat{a}_{\nu\bm{k}}, 
\end{align}
where $A^{\mu}_v(\bm{q})=\sqrt{\hbar/2m_v\Omega_{\mu}(\mathbf{q})}$ is the phonon amplitude, $\Omega_{\mu}(\bm{q})$ and $\bm{e}_t^{\mu}(\bm{q})$ are the eigenvalue and eigenstate of $\mu$-th phonon mode, $\hat{b}$ ($\hat{b}^{\dagger}$) is the phonon annihilation (creation) operator, and $f_s^{\nu}(\bm{k})$ is the eigenstate of $\nu$-th magnon mode. After straightforward calculations \cite{Delugas2023,supp}, the linear magnon-phonon coupling Hamiltonian in a second quantization form becomes
\begin{equation}
        H_{\mathrm{mag-ph}} = -\sum_{\mu\nu\bm{q}}(\lambda_{\nu\mu}(\bm{q})\hat{a}^{\dagger}_{\nu\bm{q}}\hat{b}_{\mu\bm{q}} + \lambda^*_{\nu\mu}(\bm{q})\hat{a}_{\nu\bm{q}}\hat{b}^{\dagger}_{\mu\bm{q}}), 
\end{equation}
where $\lambda_{\nu\mu}(\bm{q})$ is the linear magnon-phonon coupling matrix. It is calculated as
\begin{equation}\label{eq:coupling1}
    \lambda_{\nu\mu}(\bm{q}) = \frac{S}{2}\sqrt{2S}\sum_{s,t,v}f_s^{\nu}(-\bm{q})\mathcal{J}'_{st}(v,\bm{q})\bm{e}_v^{\mu}(\bm{q})A_v^{\mu}(\bm{q}),
\end{equation}
where we define the Fourier transform of $\nabla_{lv} \mathcal{J}_{ns,mt}^{-z}$ as 
\begin{align}\label{eq:FT}
    \mathcal{J}'_{st}(v,\bm{q}) = \sum_{\bm{L},\bm{L'}}e^{i\bm{q}\cdot\bm{L'}}\nabla_{\bm{L'}v} \mathcal{J}_{st}^{-z}(\bm{L})
\end{align}
Due to the translation symmetry, we use $\nabla_{\bm{L'}v} \mathcal{J}_{st}^{-z}(\bm{L})$ to represent $\nabla_{lv} \mathcal{J}_{ns,mt}^{-z}$, where $\bm{L}=\bm{R}_m-\bm{R}_n$ and $\bm{L'}=\bm{R}_l-\bm{R}_n$.

\noindent
Before presenting the results for real material systems, we highlight some general considerations for the practical implementation of our formalism. The derivative $\nabla_{\bm{L'}v} \mathcal{J}_{st}^{-z}(\bm{L})$ is computed in real space using a supercell approach. The supercell size should be large enough to include all neighbors of the $ns$-th magnetic atom up to a specified distance. The derivative can be computed using the finite difference method by displacing the $lv$-th atom (both magnetic and non-magnetic) along the $x$, $y$, and $z$ directions in the supercell. However, this procedure requires several hundred calculations, and the displacement step needs to be small and well converged. Moreover, obtaining accurate derivatives is also challenging given that the magnitude of the off-diagonal exchange constant is of the order of 0.1 meV and the displacement step is of the order of 0.01~\AA. We instead apply the Hellmann-Feynman force theorem to compute the forces acting on each atom that can be obtained accurately from DFT, and use these to compute the derivative. This procedure completely avoids finite difference calculations. More details can be found in the section \ref{sec:details}. Eq.~(\ref{eq:FT}) allows us to compute the Fourier transform of the derivative $\nabla_{\bm{L'}v} \mathcal{J}_{st}^{-z}(\bm{L})$ on an arbitrary vector $\bm{q}$, but it is only exact for a vector $\bm{q}$ commensurate to the size of the supercell. For an incommensurate vector $\bm{q}$ we need to perform Fourier interpolation. The interpolation is reliable when the derivative $\nabla_{\bm{L'}v} \mathcal{J}_{st}^{-z}(\bm{L})$ decays rapidly as $\lvert\bm{L'}\rvert$ increases, similar to the procedure for interpolating the dynamical matrix from force constants in phonon calculations \cite{DFPT}. For the phase factor in Eq. (\ref{eq:FT}), there is an ambiguity in choosing the origin of the supercell. Following the convention used in Ref.~\cite{phonopy-phono3py-JPCM}, we choose the supercell vector $\bm{L''}$ that gives the shortest vector $\bm{R}_l-\bm{R}_n+\bm{L''}$ as the origin.

\section{Results and discussions}

\subsection{Calculations of magnon spectra and dependence on exchange-correlation functional}
\begin{figure}[htbp]
    \centering
    \includegraphics[width=\columnwidth]{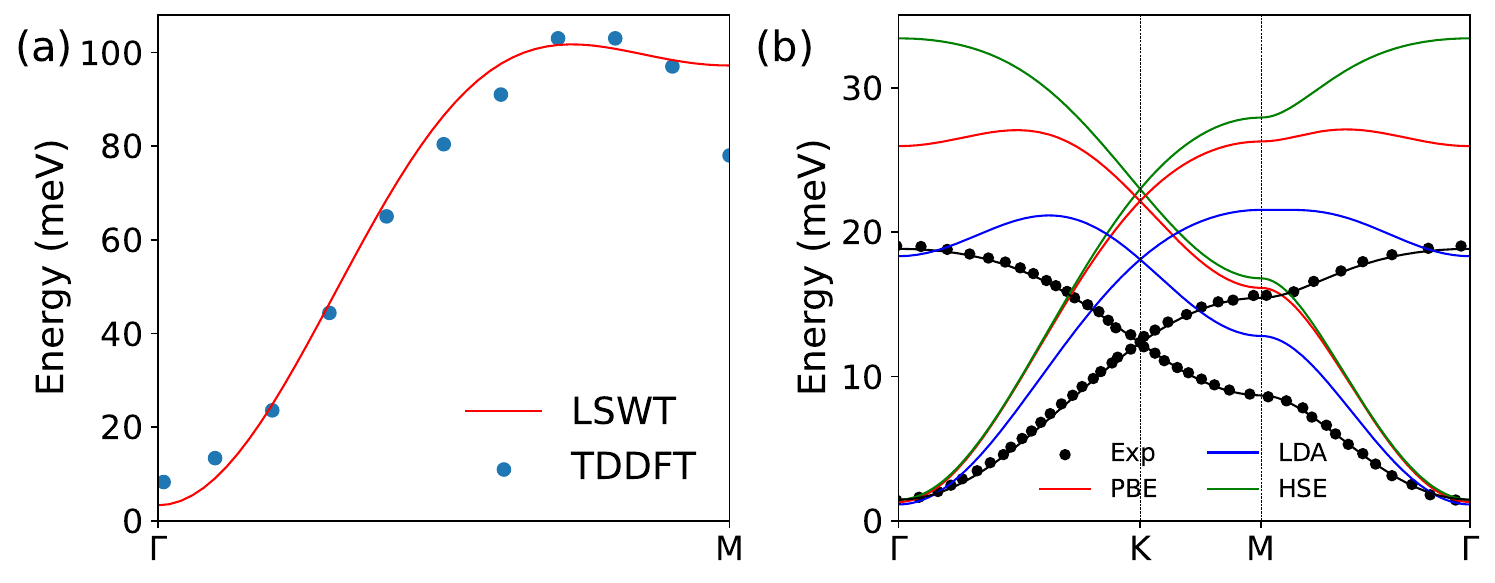}
    \caption{(a) Magnon band structure of $\mathrm{CrTe_2}$ monolayer from LSWT and TDDFT (b) Magnon band structure of $\mathrm{CrI_3}$ monolayer using LSWT with exchange and single-ion anisotropy constants from DFT with different exchange-correlation functionals. The black curve are plotted using the exchange constants (Table \ref{tab:exchange}) fitted from experiment \cite{Chen2018}.}
    \label{fig:magnon}
\end{figure}

\begin{table}[htbp]
\centering
\caption{Exchange and single-ion anisotropy constants in meV. $S=3/2$ is used.}
\begin{tabular}{ccccccccc}
    \hline
    System            & Method               & $J_1$ & $J_2$ & $J_3$ & $K$   & $\mathcal{J}_1^{xz}$ & $\mathcal{J}_1^{yz}$  \\ \hline
    $\mathrm{CrI_3}$  & DFT (LDA)            & 2.16  & 0.62  & -0.25 & 0.38  & 0.16                 & 0.28 \\ \hline
    $\mathrm{CrI_3}$  & DFT (PBE)            & 2.90  & 0.63  & -0.16 & 0.44  & 0.20                 & 0.35 \\ \hline
    $\mathrm{CrI_3}$  & DFT (HSE06)          & 3.59  & 0.41  & -0.04 &       \\ \hline
    $\mathrm{CrI_3}$  & Exp \cite{Chen2018}  & 2.01  & 0.16  & -0.08 & 0.49  \\ \hline
    $\mathrm{CrTe_2}$ & DFT (LDA)            & 4.80  & 3.02  &  1.51 & 1.11  \\ \hline
    $\mathrm{CrTe_2}$ & DFT (PBE)            & 7.46  & 3.38  &  0.62 & 1.64  \\ \hline
\end{tabular}
\label{tab:exchange}
\end{table}

The diagonal exchange and single-ion magnetic anisotropy constants for monolayer $\mathrm{CrTe_2}$ and $\mathrm{CrI_3}$ are summarized in Table~\ref{tab:exchange}. We use $J_1$, $J_2$, and $J_3$ to represent the exchange constants, the diagonal coefficient of the exchange tensor, between one Cr atom and its first, second, and third-nearest neighbors, respectively. The magnetic moment of Cr atom is calculated to be close to $3\mu_B$ from DFT calculations, so we take the spin of Cr atom to be $S=3/2$. Our calculated exchange constants for the monolayer $\mathrm{CrI_3}$ are in excellent agreement with a previous theoretical study \cite{Cong2022}. In Fig.~\ref{fig:magnon} (a), we show the magnon band structure of monolayer $\mathrm{CrTe_2}$ computed from LSWT and linear response TDDFT. The latter is calculated within the Liouville-Lanczos approach \cite{turboMagnon}, and incorporates spin-orbit coupling self-consistently. 
Due to the lack of exchange-correlation functionals for non collinear spin DFT, local density approximation (LDA) is used in both methods for a fair comparison. However, we note that the choice of exchange-correlation functional is rather flexible for LSWT through exchange constants, and we compare the results among different functionals later. Overall, the two methods show good agreement except at the zone border, which suggests that the magnon eigenstates from LSWT are comparable to those from TDDFT when we use the same exchange-correlation functional. We also note that the magnon gap at $\Gamma$ is captured in both TDDFT and LSWT calculations. 

In Fig.~\ref{fig:magnon} (b), we show for monolayer $\mathrm{CrI_3}$ the magnon band structure calculated from LSWT, using the exchange and single-ion anisotropy constants obtained from DFT using LDA \cite{LDA-CA}, PBE \cite{PBE}, and HSE06 \cite{HSE} exchange-correlation functionals. 
We also show experimental data from inelastic neutron scattering measurements \cite{Chen2018}. The experimental exchange constants in Table~\ref{tab:exchange} are fitted from the experimental data using the same Hamiltonian that we use for the LSWT calculations. Interestingly, we found that the LDA is in best agreement with the experiments. 
The disagreement between theory and experiments could be explained in terms of the inaccuracy of the exchange-correlation functional used to compute the exchange constants in monolayer $\mathrm{CrI_3}$ \cite{ke_electron_2021}. 
We will show in the next section that, though the diagonal exchange constants are sensitive to the choice of exchange-correlation functional, the off-diagonal exchange constants are not. 

\subsection{Numerical implementation of magnon-phonon coupling}

\begin{figure}[htbp]
    \centering
    \includegraphics[width=\columnwidth]{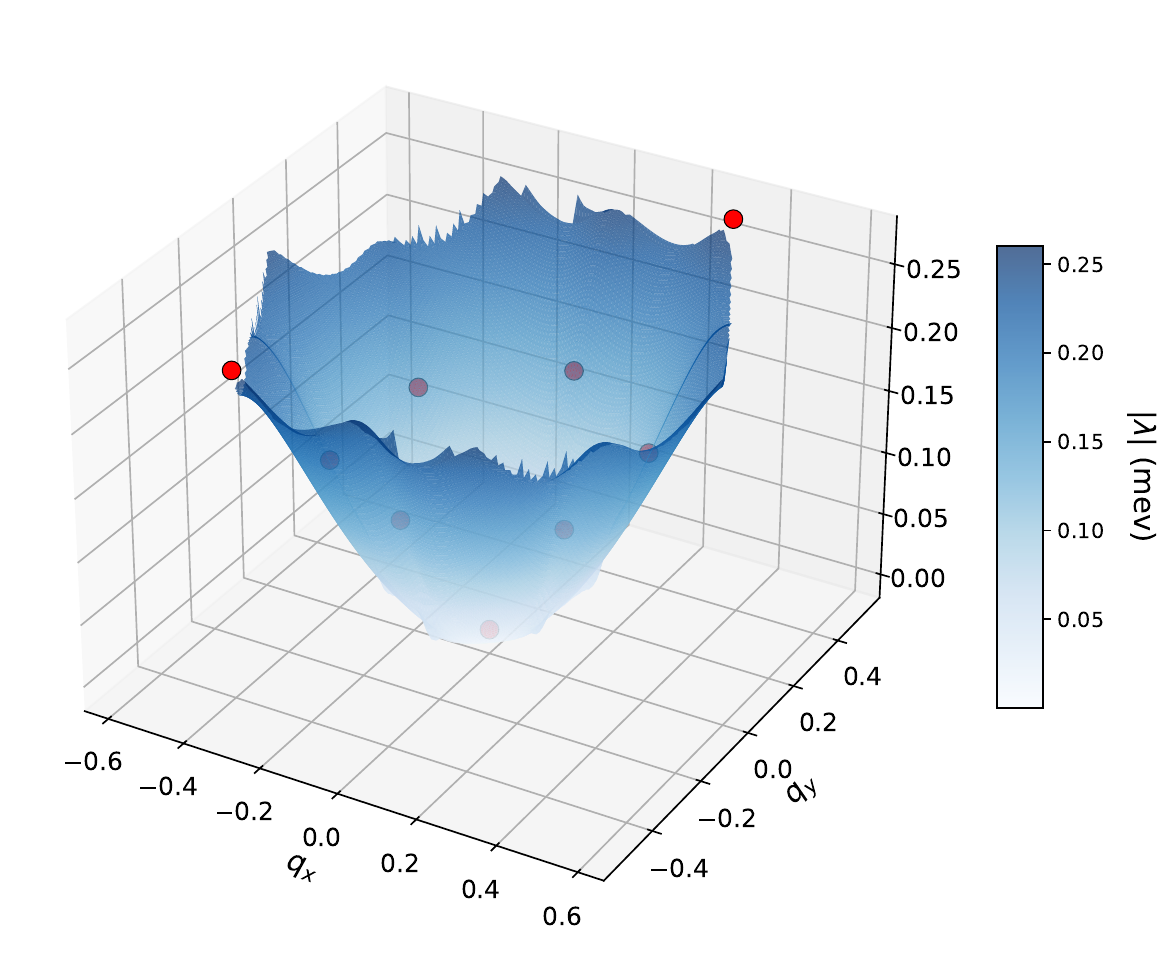}
    \caption{Linear magnon-phonon coupling constant between acoustic magnon mode and 16th phonon mode in the Brillouin zone (BZ) for monolayer $\mathrm{CrI_3}$. Blue surface: coupling constants from Fourier interpolation. Red dots: coupling constants on a $3\times3\times1$ $\bm{q}$-grids commensurate with the supercell size at $3\times3\times1$.}
    \label{fig:3d_coupling}
\end{figure}

In Fig.~\ref{fig:3d_coupling}, we show the linear magnon-phonon coupling constant between the acoustic magnon mode and the 16th phonon mode in the entire first Brillouin zone (BZ) of monolayer $\mathrm{CrI_3}$. We use a $3\times3\times1$ supercell to compute $\nabla_{lv} \mathcal{J}_{ns,mt}^{-z}$, then perform Fourier interpolation for each $\bm{q}$ vector in the BZ to get $\mathcal{J}'_{st}(v,\bm{q})$, and feed it into Eq.~(\ref{eq:coupling1}) to get the coupling constant. The $3\times3\times1$ $\bm{q}$-grids are commensurate with the supercell size so that $\mathcal{J}'_{st}(v,\bm{q})$ is the exact Fourier transform of $\nabla_{lv} \mathcal{J}_{ns,mt}^{-z}$. We see that the coupling constants on the $\bm{q}$-grids (red points) fall on the Fourier-interpolated continuous surface. This demonstrates the reliability of the interpolation procedure. 

\begin{figure}[htbp]
    \centering
    \includegraphics[width=\columnwidth]{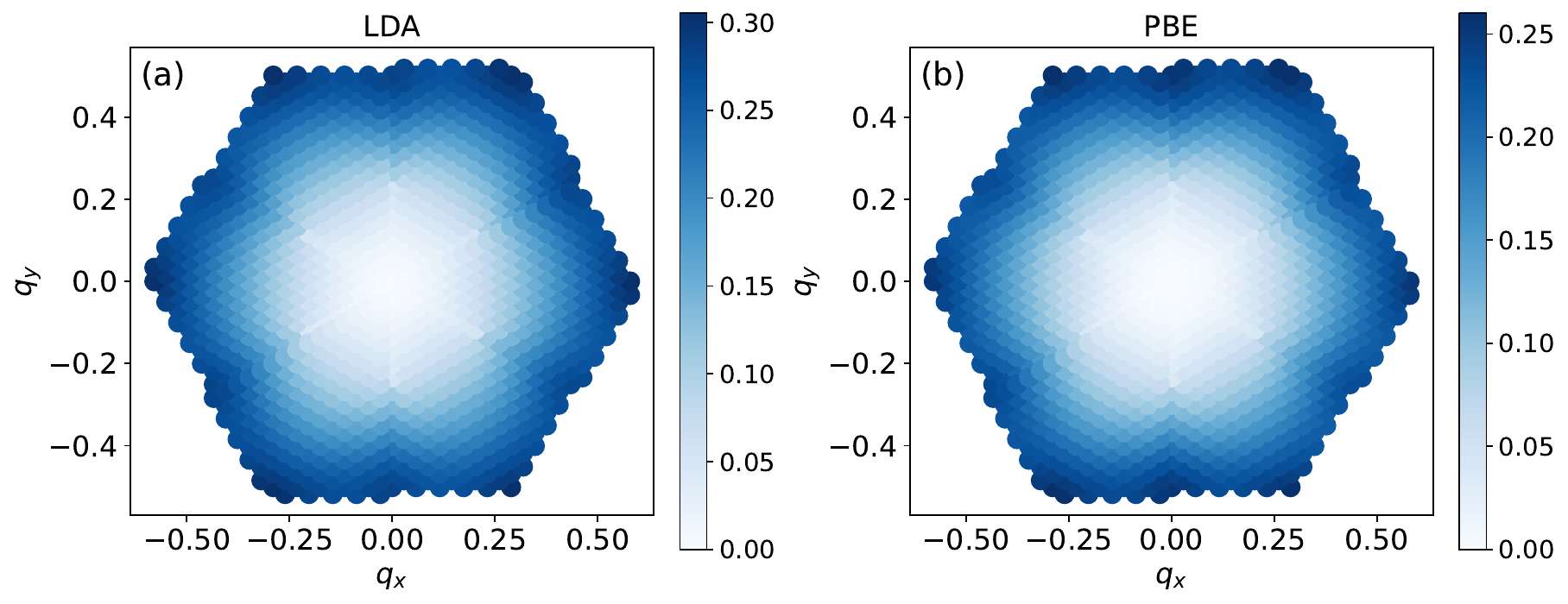}
    \caption{Linear magnon-phonon coupling constants between acoustic magnon mode and 16th phonon mode in the BZ for monolayer $\mathrm{CrI_3}$ using (a) LDA and (b) PBE exchange-correlation functional.}
    \label{fig:coupling_XC}
\end{figure}

In Fig.~\ref{fig:magnon}(b), we show that the magnon band structures, which are determined by diagonal exchange constants as shown in Eq.~(\ref{eq:Hmag}), are sensitive to the choice of exchange-correlation functional. However, the off-diagonal exchange constants and hence their derivatives are less sensitive to the choice of exchange-correlation functional, as shown in Table~\ref{tab:exchange}. In Fig.~\ref{fig:coupling_XC}, we show the linear coupling constants between the acoustic magnon mode and the 16th phonon mode in the entire Brillouin zone (BZ) for the monolayer $\mathrm{CrI_3}$ using LDA and PBE exchange-correlation functionals. Except for a minor difference in the coupling strength, the two functionals produce similar patterns. This demonstrates that the coupling constants, which are obtained from the derivatives of off-diagonal exchange constants, are less sensitive to the choice of exchange-correlation functional.

\begin{figure}[htbp]
    \centering
    \includegraphics[width=0.85\columnwidth]{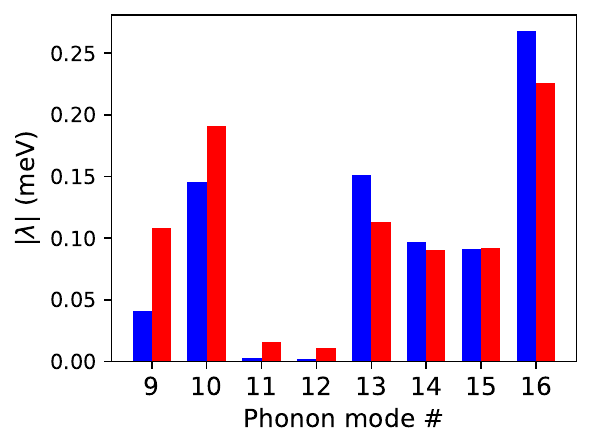}
    \caption{Linear magnon-phonon coupling constant between the magnon acoustic mode and various phonon modes in monolayer $\mathrm{CrI_3}$ at Dirac ($K$) point computed from Eq.~(\ref{eq:coupling1}) (blue) and Eq.~(\ref{eq:coupling2}) (red).}
    \label{fig:coupling_K}
\end{figure}

As shown in the previous study \cite{Delugas2023}, the linear magnon-phonon coupling can also be obtained from the stationary condition of magnon operator $\dot{\hat{a}}=\frac{i}{\hbar}[H,\hat{a}]=0$, resulting in the following equation
\begin{equation}\label{eq:coupling2}
    \lambda_{\nu\mu}(\bm{q}) = \frac{\hbar\omega_{\nu}(\bm{q})\langle\hat{a}_{\nu\bm{q}}\rangle}{\langle\hat{b}_{\mu\bm{q}}\rangle},
\end{equation}
where $\langle\ldots\rangle$ represents the expectation value of the operator. In practice, the coupling constant using Eq.~(\ref{eq:coupling2}) is calculated by distorting the lattice in a supercell along the $\mu$-th phonon mode at a certain $\bm{q}$ vector $\langle\hat{b}_{\mu\bm{q}}\rangle$, which induces a tilting of magnetization, caused by a non-zero $\langle\hat{a}_{\nu\bm{q}}\rangle$, from the $z$ axis towards the plane. This procedure has to be repeated for each phonon mode at each $\bm{q}$ vector if we want to get the coupling constant across the BZ. However, using Eq.~(\ref{eq:coupling1}) gives the coupling constant across the BZ with negligible computational cost, since the information in reciprocal space is encoded in the Fourier interpolation of $\nabla_{lv} \mathcal{J}_{ns,mt}^{-z}$, which is calculated once in real space. In Fig. \ref{fig:coupling_K}, we compare the coupling constants at Dirac (K) point for monolayer $\mathrm{CrI_3}$ computed from Eq.~(\ref{eq:coupling1}) and~(\ref{eq:coupling2}). The coefficients obtained from the two methods appear in reasonably good agreement. Specifically, the 11th and 12th phonon modes have negligible coupling with the magnon, while the 16th phonon mode has the strongest. Our results are also consistent with a previous study using Eq.~(\ref{eq:coupling2}) \cite{Delugas2023}.

\begin{figure}[htbp]
    \centering
    \includegraphics[width=\columnwidth]{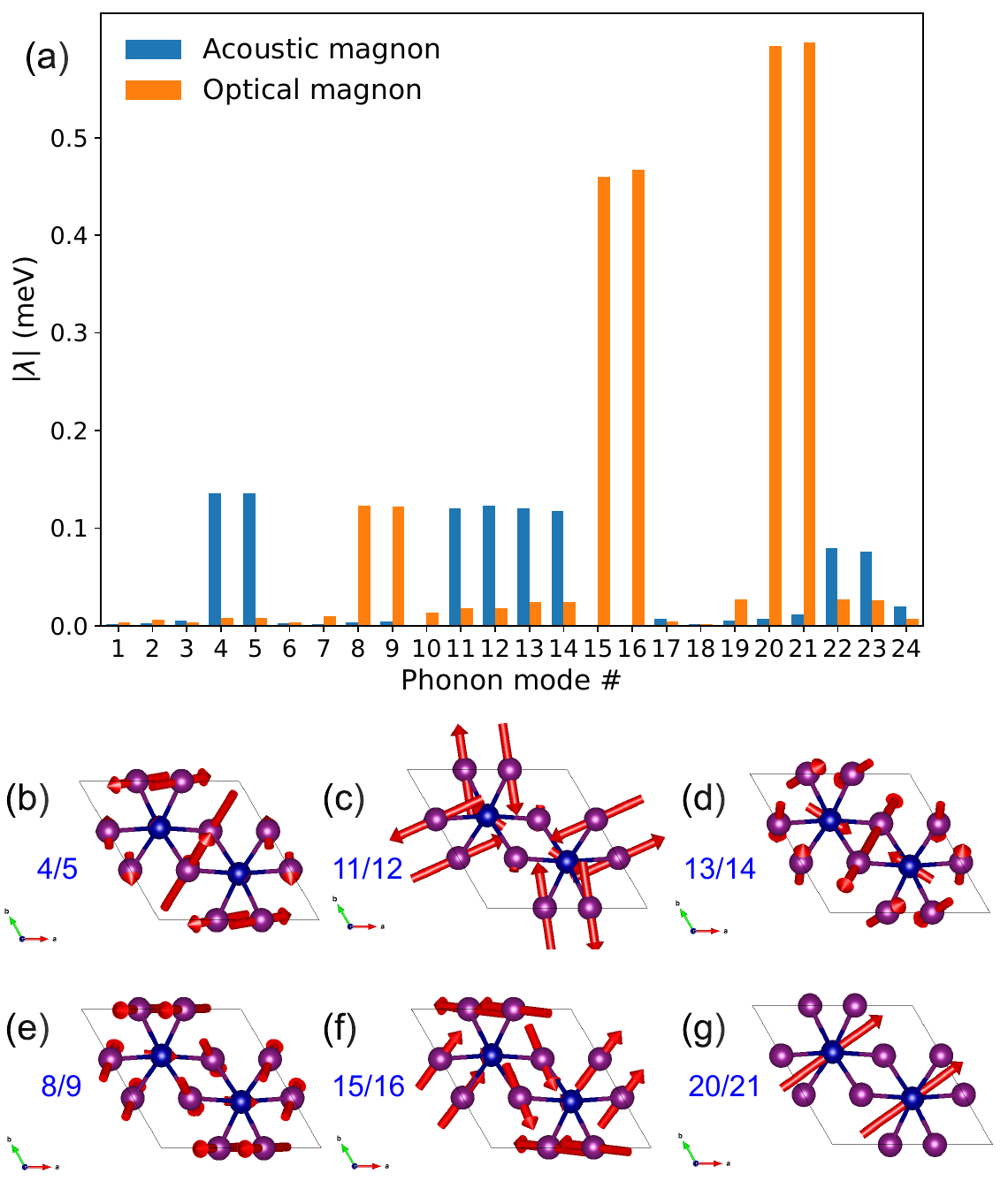}
    \caption{(a) Linear magnon-phonon coupling constant between different magnon and phonon modes in monolayer $\mathrm{CrI_3}$ at $\Gamma$ point. Phonon vibrations in (b) 4th/5th (c) 11th/12th (d) 13th/14th (e) 8th/9th (f) 15th/16th (g) 20th/21st modes. Blue atoms represent Cr atoms and purple atoms represent I atoms.}
    \label{fig:cri3_G}
\end{figure}

Eq.~(\ref{eq:coupling1}) shows that the linear magnon-phonon coupling constant is a function of $\mathcal{J}'_{st}(v,\bm{q})$, and of the magnon and phonon eigenstates, which are all dependent on $\bm{q}$ vectors. In Fig.~\ref{fig:cri3_G}(a), we show the coupling constants between the magnon and phonon modes at $\Gamma$ point in the monolayer $\mathrm{CrI_3}$. The coupling constants at the $\Gamma$ point provide valuable information given that experimental techniques such as magneto-Raman and magneto-infrared spectroscopy can only probe the magnon and phonon at $\Gamma$. At the $\Gamma$ point, the coupling constants exhibit distinct behaviors for the acoustic and optical magnon modes. In addition, we plot the phonon motions along the modes with a strong coupling with the magnons. We observe that for the acoustic magnon mode, the phonon modes that pair more strongly with the magnons involve the out-of-phase motions of Cr or I atoms, as shown in Fig.~\ref{fig:cri3_G}(b), (c), and (d). In contrast, for optical magnon mode, the phonon modes that strongly couple with magnons involve the in-phase motions of Cr or I atoms as shown in Fig.~\ref{fig:cri3_G}(e), (f), and (g). Specifically, the phonon modes as shown in Fig.~\ref{fig:cri3_G} (e), (f), and (g) do not change the relative distance between Cr atoms but all show strong couplings with magnons, which suggests the phonon motions induce a large change in off-diagonal exchange constants, resulting in the strong couplings. In the literature, when deriving magnon-phonon coupling from Hamiltonian, it is often approximated that the exchange constants depend solely on the relative distance between the magnetic atoms \cite{Streib2019,Cong2022}. This approximation is not valid if the magnetic interaction is of superexchange type as in the monolayer $\mathrm{CrI_3}$ \cite{Lado_2017}. In Table II in SI \cite{supp}, we show that the derivatives of diagonal and off-diagonal exchange constants of monolayer $\mathrm{CrI_3}$, which demonstrates that the displacement of the non-magnetic atoms has a significant impact on the both diagonal and off-diagonal exchange constants. This also suggests that it is necessary to include the phonon displacement from every atom in the supercell as shown in Eq.~(\ref{eq:mag-ph}) when deriving magnon-phonon coupling Hamiltonian.

\begin{figure}[htbp]
    \centering
    \includegraphics[width=\columnwidth]{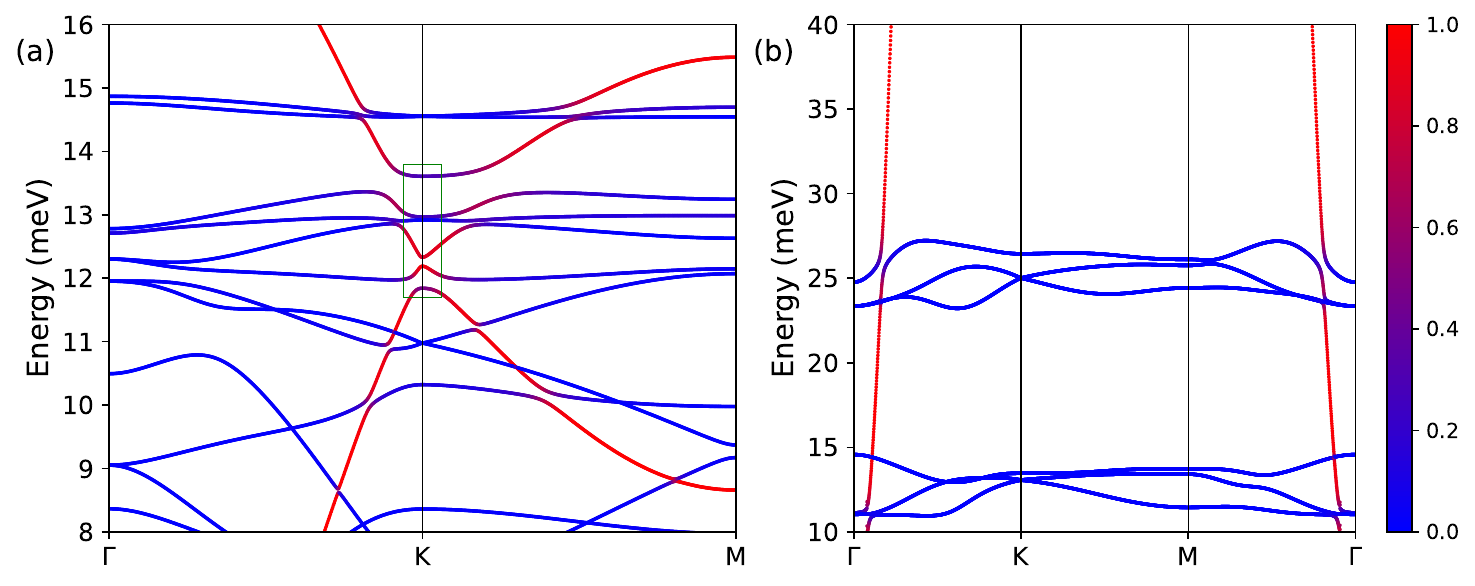}
    \caption{Hybridized magnon-phonon band structure for monolayer (a) $\mathrm{CrI_3}$ and (b) $\mathrm{CrTe_2}$. The green rectangle in (a) indicates the opening gap induced by linear magnon-phonon coupling at K point.}
    \label{fig:band}
\end{figure}

The couplings between magnons and phonons do not always manifest an anticrossing feature in the hybridized magnon-phonon band structure. In order to observe such a feature, the magnon and phonon should possess very similar momentum and energy, in addition to being strongly coupled. As shown in Fig.~\ref{fig:band}(a), the magnon at the Dirac (K) point interacts strongly with the nearby phonon modes, resulting in an gap opening in monolayer $\mathrm{CrI_3}$. Here the diagonal exchange constants from the experiment are adopted (Table~\ref{tab:exchange}), although the rest elements are from our calculations. This is because the diagonal exchange constants of this system are sensitive to the choice of exchange-corelation functional but off-diagonal ones are not as we discussed previously. The results with fully first-principles exchange constants are detailed in SI Fig.2.
The opening gap is about 2 meV as indicated in the green rectangle in Fig.~\ref{fig:band}(a), which is comparable to the 4 meV observed in experiment \cite{Chen2018}. As shown in Fig.~\ref{fig:band}(b), in monolayer $\mathrm{CrTe_2}$, the strong coupling results in an anticrossing feature near $\Gamma$ point at 25 meV but no such feature is observed at 15 meV because of the weak magnon-phonon coupling. We note that the position of crossing point on the magnon band structure is determined by diagonal exchange constants, which may be easily tuned, e.g. by applying strain. On the other hand, the coupling strength or anti-crossing gap is determined by off-diagonal elements of exchange constant tensor (Eq.~(\ref{eq:coupling1})), which may be controlled by activating a particular phonon mode e.g. through Terahertz laser, for Terahertz-phonon-magnon transduction.    

\section{Conclusion and outlook}
In conclusion, we have developed an efficient first-principles approach to compute magnon spectra and magnon-phonon coupling constants based on linear spin wave theory at an arbitrary $\bm{q}$ vector in the BZ. The coupling constants are determined by the derivative of off-diagonal exchange constants, and magnon and phonon eigenstates. The derivatives are calculated from the Hellmann-Feynman forces in a real-space supercell approach, avoiding the use of finite difference methods. Our results show that though the diagonal exchange constants are sensitive to the choice of exchange-correlation functional, the magnon-phonon coupling constants are not. 

This approach can be readily extended to calculate higher-order magnon-phonon coupling constants depending on the derivatives of the diagonal exchange constants, which can be used to study the relaxation and decoherence of quantum magnon states. Our approach provides a valuable tool for studying hybridized magnon-phonon states and tuning the coupling strength via external stimuli such as strain, doping, and Terahertz excitation, which are crucial for understanding the coupling mechanism and designing novel hybrid magnon-phonon spintronic devices.

\section{\label{sec:details}Computational details}

Density functional theory (DFT) calculations are performed using the projector augmented-wave (PAW) method \cite{Blochl1994} implemented in the Vienna Ab Initio Simulation Package (\textsc{vasp}) \cite{VASP1,VASP2}. $3p^63d^54s^1$ orbitals of Cr, $5s^25p^5$ orbitals of I, and $5s^25p^4$ orbitals of Te are treated as valence electrons. Unless otherwise specified, the generalized gradient approximation of Perdew-Burke-Ernzerhof (PBE) \cite{PBE} is used to approximate the exchange-correlation functional. Local density approximation (LDA) functional \cite{LDA-CA} and Heyd-Scuseria-Ernzehof (HSE) hybrid functional \cite{HSE} are also used to study the functional dependence. A plane-wave basis set with a kinetic energy cutoff of 500 eV is used. A $\Gamma$-centered $18\times18\times1$ for $\mathrm{CrTe_2}$ and a $8\times8\times1$ for $\mathrm{CrI_3}$ $k$-point meshes are used for integration in the first Brillouin zone (BZ). A vacuum layer of thickness 20 $\mathrm{\AA}$ is used for the monolayer (ML) structure, which is fully relaxed until the Hellmann-Feynman force on each atom is less than 1 meV/$\mathrm{\AA}$. The lattice constants are 7.00 $\mathrm{\AA}$ for $\mathrm{CrI_3}$ and 3.91 $\mathrm{\AA}$ for $\mathrm{CrTe_2}$.

The phonon properties are calculated using \textsc{phonopy} code \cite{phonopy-phono3py-JPCM,phonopy-phono3py-JPSJ}. The interatomic force constants (IFCs) are calculated based on the finite-difference method using a $3\times3\times1$ supercell for monolayer $\mathrm{CrI_3}$ and $6\times6\times1$ supercell for monolayer $\mathrm{CrTe_2}$. For monolayer $\mathrm{CrTe_2}$, phonon band structure is only stable with the on-site Coulomb interaction ($U$) on Cr $3d$ orbital, which is described within the DFT+U approach \cite{DFT+U} with $U=2$ eV. Phonon eigenstates are obtained by diagonalizing the dynamical matrix, which is built upon IFCs.

Diagonal exchange constants are calculated from a total energy mapping approach with different collinear spin states (spin quantization axis is along the $z$ direction) in a supercell. Off-diagonal exchange constant and its derivative are calculated using the four-state methodology \cite{Xiang2011} with four \emph{noncollinear} spin configurations in a supercell. For example, to calculate $\mathcal{J}_{ij}^{xz}$, the following four noncollinear spin configurations are employed: (1) $S_i=(S,0,0), S_j=(0,0,S)$; (2) $S_i=(S,0,0), S_j=(0,0,-S)$; (3) $S_i=(-S,0,0), S_j=(0,0,S)$; (4) $S_i=(-S,0,0), S_j=(0,0,-S)$ with all other spin states in the supercell set to $(0,0,S)$. $\mathcal{J}_{ij}^{xz}=(E_1-E_2-E_3+E_4)/4S^2$, where $E_{1-4}$ are the total energies of the four noncollinear spin configurations calculated from DFT. The four-state methodology also allows to compute the derivative of the exchange constant by using the Hellmann-Feynman force. To calculate $\partial \mathcal{J}_{ij}^{xz}/\partial\bm{u}_k$, where $\bm{u}_k$ is the displacement of $k$-th atom in the supercell, we use $\partial \mathcal{J}_{ij}^{xz}/\partial\bm{u}_k=(\bm{F}_1^k-\bm{F}_2^k-\bm{F}_3^k+\bm{F}_4^k)/4S^2$, where $\bm{F}^k_{1-4}$ are the Hellmann-Feynman forces of the four noncollinear spin states acting on $k$-th atom. The single-ion anisotropy constant is calculated from the difference in total energy with magnetization aligned along $a$ and $c$ axis with spin-orbit coupling turned on.

Time-dependent density functional theory (TDDFT) magnon calculations of monolayer $\mathrm{CrTe_2}$ are performed using the turboMagnon module \cite{turboMagnon} in Quantum ESPRESSO code \cite{QE1,QE2}. The optimized norm-conserving Vanderbilt (ONCV) pseudopotentials from the PseudoDojo project \cite{ONCV1, van2018pseudodojo} are used with a kinetic energy cutoff of 60 Ry. Adiabatic local density approximation (ALDA) is used as the exchange-correlation functional. A $\Gamma$-centered $24\times24\times1$ $k$-point mesh is used for integration in the BZ. Spin-orbit coupling (SOC) is included self-consistently. 30000 Lanczos steps are used to converge the magnon spectra. The convergence tests are included in Supplemental Material \cite{supp}. 

\begin{acknowledgments}
This research was primarily supported by NSF through the University of Wisconsin Materials Research Science and Engineering Center (DMR-2309000). The authors acknowledge the Texas Advanced Computing Center (TACC) at The University of Texas at Austin for providing computational resources that have contributed to the research results reported within this paper.
\end{acknowledgments}

%


\begin{thebibliography}{40}%
\makeatletter
\providecommand \@ifxundefined [1]{%
 \@ifx{#1\undefined}
}%
\providecommand \@ifnum [1]{%
 \ifnum #1\expandafter \@firstoftwo
 \else \expandafter \@secondoftwo
 \fi
}%
\providecommand \@ifx [1]{%
 \ifx #1\expandafter \@firstoftwo
 \else \expandafter \@secondoftwo
 \fi
}%
\providecommand \natexlab [1]{#1}%
\providecommand \enquote  [1]{``#1''}%
\providecommand \bibnamefont  [1]{#1}%
\providecommand \bibfnamefont [1]{#1}%
\providecommand \citenamefont [1]{#1}%
\providecommand \href@noop [0]{\@secondoftwo}%
\providecommand \href [0]{\begingroup \@sanitize@url \@href}%
\providecommand \@href[1]{\@@startlink{#1}\@@href}%
\providecommand \@@href[1]{\endgroup#1\@@endlink}%
\providecommand \@sanitize@url [0]{\catcode `\\12\catcode `\$12\catcode `\&12\catcode `\#12\catcode `\^12\catcode `\_12\catcode `\%12\relax}%
\providecommand \@@startlink[1]{}%
\providecommand \@@endlink[0]{}%
\providecommand \url  [0]{\begingroup\@sanitize@url \@url }%
\providecommand \@url [1]{\endgroup\@href {#1}{\urlprefix }}%
\providecommand \urlprefix  [0]{URL }%
\providecommand \Eprint [0]{\href }%
\providecommand \doibase [0]{https://doi.org/}%
\providecommand \selectlanguage [0]{\@gobble}%
\providecommand \bibinfo  [0]{\@secondoftwo}%
\providecommand \bibfield  [0]{\@secondoftwo}%
\providecommand \translation [1]{[#1]}%
\providecommand \BibitemOpen [0]{}%
\providecommand \bibitemStop [0]{}%
\providecommand \bibitemNoStop [0]{.\EOS\space}%
\providecommand \EOS [0]{\spacefactor3000\relax}%
\providecommand \BibitemShut  [1]{\csname bibitem#1\endcsname}%
\let\auto@bib@innerbib\@empty
\bibitem [{\citenamefont {Chumak}\ \emph {et~al.}(2015)\citenamefont {Chumak}, \citenamefont {Vasyuchka}, \citenamefont {Serga},\ and\ \citenamefont {Hillebrands}}]{chumak_magnon_2015}%
  \BibitemOpen
  \bibfield  {author} {\bibinfo {author} {\bibfnamefont {A.~V.}\ \bibnamefont {Chumak}}, \bibinfo {author} {\bibfnamefont {V.~I.}\ \bibnamefont {Vasyuchka}}, \bibinfo {author} {\bibfnamefont {A.~A.}\ \bibnamefont {Serga}},\ and\ \bibinfo {author} {\bibfnamefont {B.}~\bibnamefont {Hillebrands}},\ }\bibfield  {title} {\bibinfo {title} {Magnon spintronics},\ }\href {https://doi.org/10.1038/nphys3347} {\bibfield  {journal} {\bibinfo  {journal} {Nature Physics}\ }\textbf {\bibinfo {volume} {11}},\ \bibinfo {pages} {453} (\bibinfo {year} {2015})}\BibitemShut {NoStop}%
\bibitem [{\citenamefont {Yuan}\ \emph {et~al.}(2022)\citenamefont {Yuan}, \citenamefont {Cao}, \citenamefont {Kamra}, \citenamefont {Duine},\ and\ \citenamefont {Yan}}]{YUAN20221}%
  \BibitemOpen
  \bibfield  {author} {\bibinfo {author} {\bibfnamefont {H.}~\bibnamefont {Yuan}}, \bibinfo {author} {\bibfnamefont {Y.}~\bibnamefont {Cao}}, \bibinfo {author} {\bibfnamefont {A.}~\bibnamefont {Kamra}}, \bibinfo {author} {\bibfnamefont {R.~A.}\ \bibnamefont {Duine}},\ and\ \bibinfo {author} {\bibfnamefont {P.}~\bibnamefont {Yan}},\ }\bibfield  {title} {\bibinfo {title} {Quantum magnonics: When magnon spintronics meets quantum information science},\ }\href {https://doi.org/https://doi.org/10.1016/j.physrep.2022.03.002} {\bibfield  {journal} {\bibinfo  {journal} {Physics Reports}\ }\textbf {\bibinfo {volume} {965}},\ \bibinfo {pages} {1} (\bibinfo {year} {2022})}\BibitemShut {NoStop}%
\bibitem [{\citenamefont {Streib}\ \emph {et~al.}(2019)\citenamefont {Streib}, \citenamefont {Vidal-Silva}, \citenamefont {Shen},\ and\ \citenamefont {Bauer}}]{Streib2019}%
  \BibitemOpen
  \bibfield  {author} {\bibinfo {author} {\bibfnamefont {S.}~\bibnamefont {Streib}}, \bibinfo {author} {\bibfnamefont {N.}~\bibnamefont {Vidal-Silva}}, \bibinfo {author} {\bibfnamefont {K.}~\bibnamefont {Shen}},\ and\ \bibinfo {author} {\bibfnamefont {G.~E.~W.}\ \bibnamefont {Bauer}},\ }\bibfield  {title} {\bibinfo {title} {Magnon-phonon interactions in magnetic insulators},\ }\href {https://doi.org/10.1103/PhysRevB.99.184442} {\bibfield  {journal} {\bibinfo  {journal} {Phys. Rev. B}\ }\textbf {\bibinfo {volume} {99}},\ \bibinfo {pages} {184442} (\bibinfo {year} {2019})}\BibitemShut {NoStop}%
\bibitem [{\citenamefont {Cong}\ \emph {et~al.}(2022)\citenamefont {Cong}, \citenamefont {Liu}, \citenamefont {Xue}, \citenamefont {Liu}, \citenamefont {Liu},\ and\ \citenamefont {Shen}}]{Cong2022}%
  \BibitemOpen
  \bibfield  {author} {\bibinfo {author} {\bibfnamefont {A.}~\bibnamefont {Cong}}, \bibinfo {author} {\bibfnamefont {J.}~\bibnamefont {Liu}}, \bibinfo {author} {\bibfnamefont {W.}~\bibnamefont {Xue}}, \bibinfo {author} {\bibfnamefont {H.}~\bibnamefont {Liu}}, \bibinfo {author} {\bibfnamefont {Y.}~\bibnamefont {Liu}},\ and\ \bibinfo {author} {\bibfnamefont {K.}~\bibnamefont {Shen}},\ }\bibfield  {title} {\bibinfo {title} {{Exchange-mediated magnon-phonon scattering in monolayer ${\mathrm{CrI}}_{3}$}},\ }\href {https://doi.org/10.1103/PhysRevB.106.214424} {\bibfield  {journal} {\bibinfo  {journal} {Phys. Rev. B}\ }\textbf {\bibinfo {volume} {106}},\ \bibinfo {pages} {214424} (\bibinfo {year} {2022})}\BibitemShut {NoStop}%
\bibitem [{\citenamefont {Berk}\ \emph {et~al.}(2019)\citenamefont {Berk}, \citenamefont {Jaris}, \citenamefont {Yang}, \citenamefont {Dhuey}, \citenamefont {Cabrini},\ and\ \citenamefont {Schmidt}}]{berk_strongly_2019}%
  \BibitemOpen
  \bibfield  {author} {\bibinfo {author} {\bibfnamefont {C.}~\bibnamefont {Berk}}, \bibinfo {author} {\bibfnamefont {M.}~\bibnamefont {Jaris}}, \bibinfo {author} {\bibfnamefont {W.}~\bibnamefont {Yang}}, \bibinfo {author} {\bibfnamefont {S.}~\bibnamefont {Dhuey}}, \bibinfo {author} {\bibfnamefont {S.}~\bibnamefont {Cabrini}},\ and\ \bibinfo {author} {\bibfnamefont {H.}~\bibnamefont {Schmidt}},\ }\bibfield  {title} {\bibinfo {title} {Strongly coupled magnon–phonon dynamics in a single nanomagnet},\ }\href {https://doi.org/10.1038/s41467-019-10545-x} {\bibfield  {journal} {\bibinfo  {journal} {Nature Communications}\ }\textbf {\bibinfo {volume} {10}},\ \bibinfo {pages} {2652} (\bibinfo {year} {2019})}\BibitemShut {NoStop}%
\bibitem [{\citenamefont {Liu}\ \emph {et~al.}(2021)\citenamefont {Liu}, \citenamefont {Granados~del \'Aguila}, \citenamefont {Bhowmick}, \citenamefont {Gan}, \citenamefont {Thu Ha~Do}, \citenamefont {Prosnikov}, \citenamefont {Sedmidubsk\'y}, \citenamefont {Sofer}, \citenamefont {Christianen}, \citenamefont {Sengupta},\ and\ \citenamefont {Xiong}}]{Liu_magnon-phonon_2021}%
  \BibitemOpen
  \bibfield  {author} {\bibinfo {author} {\bibfnamefont {S.}~\bibnamefont {Liu}}, \bibinfo {author} {\bibfnamefont {A.}~\bibnamefont {Granados~del \'Aguila}}, \bibinfo {author} {\bibfnamefont {D.}~\bibnamefont {Bhowmick}}, \bibinfo {author} {\bibfnamefont {C.~K.}\ \bibnamefont {Gan}}, \bibinfo {author} {\bibfnamefont {T.}~\bibnamefont {Thu Ha~Do}}, \bibinfo {author} {\bibfnamefont {M.~A.}\ \bibnamefont {Prosnikov}}, \bibinfo {author} {\bibfnamefont {D.}~\bibnamefont {Sedmidubsk\'y}}, \bibinfo {author} {\bibfnamefont {Z.}~\bibnamefont {Sofer}}, \bibinfo {author} {\bibfnamefont {P.~C.~M.}\ \bibnamefont {Christianen}}, \bibinfo {author} {\bibfnamefont {P.}~\bibnamefont {Sengupta}},\ and\ \bibinfo {author} {\bibfnamefont {Q.}~\bibnamefont {Xiong}},\ }\bibfield  {title} {\bibinfo {title} {Direct observation of magnon-phonon strong coupling in two-dimensional antiferromagnet at high magnetic fields},\ }\href {https://doi.org/10.1103/PhysRevLett.127.097401} {\bibfield  {journal} {\bibinfo  {journal} {Phys. Rev.
  Lett.}\ }\textbf {\bibinfo {volume} {127}},\ \bibinfo {pages} {097401} (\bibinfo {year} {2021})}\BibitemShut {NoStop}%
\bibitem [{\citenamefont {Vaclavkova}\ \emph {et~al.}(2021)\citenamefont {Vaclavkova}, \citenamefont {Palit}, \citenamefont {Wyzula}, \citenamefont {Ghosh}, \citenamefont {Delhomme}, \citenamefont {Maity}, \citenamefont {Kapuscinski}, \citenamefont {Ghosh}, \citenamefont {Veis}, \citenamefont {Grzeszczyk}, \citenamefont {Faugeras}, \citenamefont {Orlita}, \citenamefont {Datta},\ and\ \citenamefont {Potemski}}]{Vaclavkova2021}%
  \BibitemOpen
  \bibfield  {author} {\bibinfo {author} {\bibfnamefont {D.}~\bibnamefont {Vaclavkova}}, \bibinfo {author} {\bibfnamefont {M.}~\bibnamefont {Palit}}, \bibinfo {author} {\bibfnamefont {J.}~\bibnamefont {Wyzula}}, \bibinfo {author} {\bibfnamefont {S.}~\bibnamefont {Ghosh}}, \bibinfo {author} {\bibfnamefont {A.}~\bibnamefont {Delhomme}}, \bibinfo {author} {\bibfnamefont {S.}~\bibnamefont {Maity}}, \bibinfo {author} {\bibfnamefont {P.}~\bibnamefont {Kapuscinski}}, \bibinfo {author} {\bibfnamefont {A.}~\bibnamefont {Ghosh}}, \bibinfo {author} {\bibfnamefont {M.}~\bibnamefont {Veis}}, \bibinfo {author} {\bibfnamefont {M.}~\bibnamefont {Grzeszczyk}}, \bibinfo {author} {\bibfnamefont {C.}~\bibnamefont {Faugeras}}, \bibinfo {author} {\bibfnamefont {M.}~\bibnamefont {Orlita}}, \bibinfo {author} {\bibfnamefont {S.}~\bibnamefont {Datta}},\ and\ \bibinfo {author} {\bibfnamefont {M.}~\bibnamefont {Potemski}},\ }\bibfield  {title} {\bibinfo {title} {Magnon polarons in the van der waals antiferromagnet
  $\mathrm{Fe}{\mathrm{ps}}_{3}$},\ }\href {https://doi.org/10.1103/PhysRevB.104.134437} {\bibfield  {journal} {\bibinfo  {journal} {Phys. Rev. B}\ }\textbf {\bibinfo {volume} {104}},\ \bibinfo {pages} {134437} (\bibinfo {year} {2021})}\BibitemShut {NoStop}%
\bibitem [{\citenamefont {Cui}\ \emph {et~al.}(2023)\citenamefont {Cui}, \citenamefont {Boström}, \citenamefont {Ozerov}, \citenamefont {Wu}, \citenamefont {Jiang}, \citenamefont {Chu}, \citenamefont {Li}, \citenamefont {Liu}, \citenamefont {Xu}, \citenamefont {Rubio},\ and\ \citenamefont {Zhang}}]{cui_chirality_2023}%
  \BibitemOpen
  \bibfield  {author} {\bibinfo {author} {\bibfnamefont {J.}~\bibnamefont {Cui}}, \bibinfo {author} {\bibfnamefont {E.~V.}\ \bibnamefont {Boström}}, \bibinfo {author} {\bibfnamefont {M.}~\bibnamefont {Ozerov}}, \bibinfo {author} {\bibfnamefont {F.}~\bibnamefont {Wu}}, \bibinfo {author} {\bibfnamefont {Q.}~\bibnamefont {Jiang}}, \bibinfo {author} {\bibfnamefont {J.-H.}\ \bibnamefont {Chu}}, \bibinfo {author} {\bibfnamefont {C.}~\bibnamefont {Li}}, \bibinfo {author} {\bibfnamefont {F.}~\bibnamefont {Liu}}, \bibinfo {author} {\bibfnamefont {X.}~\bibnamefont {Xu}}, \bibinfo {author} {\bibfnamefont {A.}~\bibnamefont {Rubio}},\ and\ \bibinfo {author} {\bibfnamefont {Q.}~\bibnamefont {Zhang}},\ }\bibfield  {title} {\bibinfo {title} {Chirality selective magnon-phonon hybridization and magnon-induced chiral phonons in a layered zigzag antiferromagnet},\ }\href {https://doi.org/10.1038/s41467-023-39123-y} {\bibfield  {journal} {\bibinfo  {journal} {Nature Communications}\ }\textbf {\bibinfo {volume} {14}},\ \bibinfo
  {pages} {3396} (\bibinfo {year} {2023})}\BibitemShut {NoStop}%
\bibitem [{\citenamefont {Luo}\ \emph {et~al.}(2023)\citenamefont {Luo}, \citenamefont {Li}, \citenamefont {Ye}, \citenamefont {Xu}, \citenamefont {Yan}, \citenamefont {Zhang}, \citenamefont {Ye}, \citenamefont {Chen}, \citenamefont {Hu}, \citenamefont {Teng}, \citenamefont {Smith}, \citenamefont {Yakobson}, \citenamefont {Dai}, \citenamefont {Nevidomskyy}, \citenamefont {He},\ and\ \citenamefont {Zhu}}]{luo_evidence_2023}%
  \BibitemOpen
  \bibfield  {author} {\bibinfo {author} {\bibfnamefont {J.}~\bibnamefont {Luo}}, \bibinfo {author} {\bibfnamefont {S.}~\bibnamefont {Li}}, \bibinfo {author} {\bibfnamefont {Z.}~\bibnamefont {Ye}}, \bibinfo {author} {\bibfnamefont {R.}~\bibnamefont {Xu}}, \bibinfo {author} {\bibfnamefont {H.}~\bibnamefont {Yan}}, \bibinfo {author} {\bibfnamefont {J.}~\bibnamefont {Zhang}}, \bibinfo {author} {\bibfnamefont {G.}~\bibnamefont {Ye}}, \bibinfo {author} {\bibfnamefont {L.}~\bibnamefont {Chen}}, \bibinfo {author} {\bibfnamefont {D.}~\bibnamefont {Hu}}, \bibinfo {author} {\bibfnamefont {X.}~\bibnamefont {Teng}}, \bibinfo {author} {\bibfnamefont {W.~A.}\ \bibnamefont {Smith}}, \bibinfo {author} {\bibfnamefont {B.~I.}\ \bibnamefont {Yakobson}}, \bibinfo {author} {\bibfnamefont {P.}~\bibnamefont {Dai}}, \bibinfo {author} {\bibfnamefont {A.~H.}\ \bibnamefont {Nevidomskyy}}, \bibinfo {author} {\bibfnamefont {R.}~\bibnamefont {He}},\ and\ \bibinfo {author} {\bibfnamefont {H.}~\bibnamefont {Zhu}},\ }\bibfield  {title}
  {\bibinfo {title} {Evidence for {Topological} {Magnon}–{Phonon} {Hybridization} in a {2D} {Antiferromagnet} down to the {Monolayer} {Limit}},\ }\href {https://doi.org/10.1021/acs.nanolett.3c00351} {\bibfield  {journal} {\bibinfo  {journal} {Nano Letters}\ }\textbf {\bibinfo {volume} {23}},\ \bibinfo {pages} {2023} (\bibinfo {year} {2023})}\BibitemShut {NoStop}%
\bibitem [{\citenamefont {Mai}\ \emph {et~al.}(2021)\citenamefont {Mai}, \citenamefont {Garrity}, \citenamefont {McCreary}, \citenamefont {Argo}, \citenamefont {Simpson}, \citenamefont {Doan-Nguyen}, \citenamefont {Aguilar},\ and\ \citenamefont {Walker}}]{mai_magnon-phonon_2021}%
  \BibitemOpen
  \bibfield  {author} {\bibinfo {author} {\bibfnamefont {T.~T.}\ \bibnamefont {Mai}}, \bibinfo {author} {\bibfnamefont {K.~F.}\ \bibnamefont {Garrity}}, \bibinfo {author} {\bibfnamefont {A.}~\bibnamefont {McCreary}}, \bibinfo {author} {\bibfnamefont {J.}~\bibnamefont {Argo}}, \bibinfo {author} {\bibfnamefont {J.~R.}\ \bibnamefont {Simpson}}, \bibinfo {author} {\bibfnamefont {V.}~\bibnamefont {Doan-Nguyen}}, \bibinfo {author} {\bibfnamefont {R.~V.}\ \bibnamefont {Aguilar}},\ and\ \bibinfo {author} {\bibfnamefont {A.~R.~H.}\ \bibnamefont {Walker}},\ }\bibfield  {title} {\bibinfo {title} {Magnon-phonon hybridization in {2D} antiferromagnet {MnPSe3}},\ }\href {https://doi.org/10.1126/sciadv.abj3106} {\bibfield  {journal} {\bibinfo  {journal} {Science Advances}\ }\textbf {\bibinfo {volume} {7}},\ \bibinfo {pages} {eabj3106} (\bibinfo {year} {2021})}\BibitemShut {NoStop}%
\bibitem [{\citenamefont {Li}\ \emph {et~al.}(2020)\citenamefont {Li}, \citenamefont {Zhang}, \citenamefont {Tyberkevych}, \citenamefont {Kwok}, \citenamefont {Hoffmann},\ and\ \citenamefont {Novosad}}]{li_hybrid_2020}%
  \BibitemOpen
  \bibfield  {author} {\bibinfo {author} {\bibfnamefont {Y.}~\bibnamefont {Li}}, \bibinfo {author} {\bibfnamefont {W.}~\bibnamefont {Zhang}}, \bibinfo {author} {\bibfnamefont {V.}~\bibnamefont {Tyberkevych}}, \bibinfo {author} {\bibfnamefont {W.-K.}\ \bibnamefont {Kwok}}, \bibinfo {author} {\bibfnamefont {A.}~\bibnamefont {Hoffmann}},\ and\ \bibinfo {author} {\bibfnamefont {V.}~\bibnamefont {Novosad}},\ }\bibfield  {title} {\bibinfo {title} {Hybrid magnonics: {Physics}, circuits, and applications for coherent information processing},\ }\href {https://doi.org/10.1063/5.0020277} {\bibfield  {journal} {\bibinfo  {journal} {Journal of Applied Physics}\ }\textbf {\bibinfo {volume} {128}},\ \bibinfo {pages} {130902} (\bibinfo {year} {2020})}\BibitemShut {NoStop}%
\bibitem [{\citenamefont {Hu}(2024)}]{hu_design_2024}%
  \BibitemOpen
  \bibfield  {author} {\bibinfo {author} {\bibfnamefont {J.-M.}\ \bibnamefont {Hu}},\ }\bibfield  {title} {\bibinfo {title} {Design of new-concept magnetomechanical devices by phase-field simulations},\ }\href {https://doi.org/10.1557/s43577-024-00699-5} {\bibfield  {journal} {\bibinfo  {journal} {MRS Bulletin}\ }\textbf {\bibinfo {volume} {49}},\ \bibinfo {pages} {636} (\bibinfo {year} {2024})}\BibitemShut {NoStop}%
\bibitem [{\citenamefont {Liao}\ \emph {et~al.}(2024)\citenamefont {Liao}, \citenamefont {Liu}, \citenamefont {Puebla}, \citenamefont {Shao},\ and\ \citenamefont {Otani}}]{liao_hybrid_2024}%
  \BibitemOpen
  \bibfield  {author} {\bibinfo {author} {\bibfnamefont {L.}~\bibnamefont {Liao}}, \bibinfo {author} {\bibfnamefont {J.}~\bibnamefont {Liu}}, \bibinfo {author} {\bibfnamefont {J.}~\bibnamefont {Puebla}}, \bibinfo {author} {\bibfnamefont {Q.}~\bibnamefont {Shao}},\ and\ \bibinfo {author} {\bibfnamefont {Y.}~\bibnamefont {Otani}},\ }\bibfield  {title} {\bibinfo {title} {Hybrid magnon-phonon crystals},\ }\href {https://doi.org/10.1038/s44306-024-00052-1} {\bibfield  {journal} {\bibinfo  {journal} {npj Spintronics}\ }\textbf {\bibinfo {volume} {2}},\ \bibinfo {pages} {1} (\bibinfo {year} {2024})}\BibitemShut {NoStop}%
\bibitem [{\citenamefont {Kittel}(1949)}]{Kittel_RMP}%
  \BibitemOpen
  \bibfield  {author} {\bibinfo {author} {\bibfnamefont {C.}~\bibnamefont {Kittel}},\ }\bibfield  {title} {\bibinfo {title} {Physical theory of ferromagnetic domains},\ }\href {https://doi.org/10.1103/RevModPhys.21.541} {\bibfield  {journal} {\bibinfo  {journal} {Rev. Mod. Phys.}\ }\textbf {\bibinfo {volume} {21}},\ \bibinfo {pages} {541} (\bibinfo {year} {1949})}\BibitemShut {NoStop}%
\bibitem [{\citenamefont {Delugas}\ \emph {et~al.}(2023)\citenamefont {Delugas}, \citenamefont {Baseggio}, \citenamefont {Timrov}, \citenamefont {Baroni},\ and\ \citenamefont {Gorni}}]{Delugas2023}%
  \BibitemOpen
  \bibfield  {author} {\bibinfo {author} {\bibfnamefont {P.}~\bibnamefont {Delugas}}, \bibinfo {author} {\bibfnamefont {O.}~\bibnamefont {Baseggio}}, \bibinfo {author} {\bibfnamefont {I.}~\bibnamefont {Timrov}}, \bibinfo {author} {\bibfnamefont {S.}~\bibnamefont {Baroni}},\ and\ \bibinfo {author} {\bibfnamefont {T.}~\bibnamefont {Gorni}},\ }\bibfield  {title} {\bibinfo {title} {{Magnon-phonon interactions enhance the gap at the Dirac point in the spin-wave spectra of ${\mathrm{CrI}}_{3}$ two-dimensional magnets}},\ }\href {https://doi.org/10.1103/PhysRevB.107.214452} {\bibfield  {journal} {\bibinfo  {journal} {Phys. Rev. B}\ }\textbf {\bibinfo {volume} {107}},\ \bibinfo {pages} {214452} (\bibinfo {year} {2023})}\BibitemShut {NoStop}%
\bibitem [{\citenamefont {Huang}\ \emph {et~al.}(2017)\citenamefont {Huang}, \citenamefont {Clark}, \citenamefont {Navarro-Moratalla}, \citenamefont {Klein}, \citenamefont {Cheng}, \citenamefont {Seyler}, \citenamefont {Zhong}, \citenamefont {Schmidgall}, \citenamefont {McGuire}, \citenamefont {Cobden}, \citenamefont {Yao}, \citenamefont {Xiao}, \citenamefont {Jarillo-Herrero},\ and\ \citenamefont {Xu}}]{huang_layer-dependent_2017}%
  \BibitemOpen
  \bibfield  {author} {\bibinfo {author} {\bibfnamefont {B.}~\bibnamefont {Huang}}, \bibinfo {author} {\bibfnamefont {G.}~\bibnamefont {Clark}}, \bibinfo {author} {\bibfnamefont {E.}~\bibnamefont {Navarro-Moratalla}}, \bibinfo {author} {\bibfnamefont {D.~R.}\ \bibnamefont {Klein}}, \bibinfo {author} {\bibfnamefont {R.}~\bibnamefont {Cheng}}, \bibinfo {author} {\bibfnamefont {K.~L.}\ \bibnamefont {Seyler}}, \bibinfo {author} {\bibfnamefont {D.}~\bibnamefont {Zhong}}, \bibinfo {author} {\bibfnamefont {E.}~\bibnamefont {Schmidgall}}, \bibinfo {author} {\bibfnamefont {M.~A.}\ \bibnamefont {McGuire}}, \bibinfo {author} {\bibfnamefont {D.~H.}\ \bibnamefont {Cobden}}, \bibinfo {author} {\bibfnamefont {W.}~\bibnamefont {Yao}}, \bibinfo {author} {\bibfnamefont {D.}~\bibnamefont {Xiao}}, \bibinfo {author} {\bibfnamefont {P.}~\bibnamefont {Jarillo-Herrero}},\ and\ \bibinfo {author} {\bibfnamefont {X.}~\bibnamefont {Xu}},\ }\bibfield  {title} {\bibinfo {title} {Layer-dependent ferromagnetism in a van der {Waals} crystal
  down to the monolayer limit},\ }\href {https://doi.org/10.1038/nature22391} {\bibfield  {journal} {\bibinfo  {journal} {Nature}\ }\textbf {\bibinfo {volume} {546}},\ \bibinfo {pages} {270} (\bibinfo {year} {2017})}\BibitemShut {NoStop}%
\bibitem [{\citenamefont {Zhang}\ \emph {et~al.}(2021)\citenamefont {Zhang}, \citenamefont {Lu}, \citenamefont {Liu}, \citenamefont {Niu}, \citenamefont {Sun}, \citenamefont {Cook}, \citenamefont {Vaninger}, \citenamefont {Miceli}, \citenamefont {Singh}, \citenamefont {Lian}, \citenamefont {Chang}, \citenamefont {He}, \citenamefont {Du}, \citenamefont {He}, \citenamefont {Zhang}, \citenamefont {Bian},\ and\ \citenamefont {Xu}}]{zhang_room-temperature_2021}%
  \BibitemOpen
  \bibfield  {author} {\bibinfo {author} {\bibfnamefont {X.}~\bibnamefont {Zhang}}, \bibinfo {author} {\bibfnamefont {Q.}~\bibnamefont {Lu}}, \bibinfo {author} {\bibfnamefont {W.}~\bibnamefont {Liu}}, \bibinfo {author} {\bibfnamefont {W.}~\bibnamefont {Niu}}, \bibinfo {author} {\bibfnamefont {J.}~\bibnamefont {Sun}}, \bibinfo {author} {\bibfnamefont {J.}~\bibnamefont {Cook}}, \bibinfo {author} {\bibfnamefont {M.}~\bibnamefont {Vaninger}}, \bibinfo {author} {\bibfnamefont {P.~F.}\ \bibnamefont {Miceli}}, \bibinfo {author} {\bibfnamefont {D.~J.}\ \bibnamefont {Singh}}, \bibinfo {author} {\bibfnamefont {S.-W.}\ \bibnamefont {Lian}}, \bibinfo {author} {\bibfnamefont {T.-R.}\ \bibnamefont {Chang}}, \bibinfo {author} {\bibfnamefont {X.}~\bibnamefont {He}}, \bibinfo {author} {\bibfnamefont {J.}~\bibnamefont {Du}}, \bibinfo {author} {\bibfnamefont {L.}~\bibnamefont {He}}, \bibinfo {author} {\bibfnamefont {R.}~\bibnamefont {Zhang}}, \bibinfo {author} {\bibfnamefont {G.}~\bibnamefont {Bian}},\ and\ \bibinfo {author}
  {\bibfnamefont {Y.}~\bibnamefont {Xu}},\ }\bibfield  {title} {\bibinfo {title} {Room-temperature intrinsic ferromagnetism in epitaxial {CrTe2} ultrathin films},\ }\href {https://doi.org/10.1038/s41467-021-22777-x} {\bibfield  {journal} {\bibinfo  {journal} {Nature Communications}\ }\textbf {\bibinfo {volume} {12}},\ \bibinfo {pages} {2492} (\bibinfo {year} {2021})}\BibitemShut {NoStop}%
\bibitem [{\citenamefont {Xiang}\ \emph {et~al.}(2013)\citenamefont {Xiang}, \citenamefont {Lee}, \citenamefont {Koo}, \citenamefont {Gong},\ and\ \citenamefont {Whangbo}}]{Xiang2013}%
  \BibitemOpen
  \bibfield  {author} {\bibinfo {author} {\bibfnamefont {H.}~\bibnamefont {Xiang}}, \bibinfo {author} {\bibfnamefont {C.}~\bibnamefont {Lee}}, \bibinfo {author} {\bibfnamefont {H.-J.}\ \bibnamefont {Koo}}, \bibinfo {author} {\bibfnamefont {X.}~\bibnamefont {Gong}},\ and\ \bibinfo {author} {\bibfnamefont {M.-H.}\ \bibnamefont {Whangbo}},\ }\bibfield  {title} {\bibinfo {title} {Magnetic properties and energy-mapping analysis},\ }\href {https://doi.org/10.1039/C2DT31662E} {\bibfield  {journal} {\bibinfo  {journal} {Dalton Trans.}\ }\textbf {\bibinfo {volume} {42}},\ \bibinfo {pages} {823} (\bibinfo {year} {2013})}\BibitemShut {NoStop}%
\bibitem [{\citenamefont {Toth}\ and\ \citenamefont {Lake}(2015)}]{Toth_2015}%
  \BibitemOpen
  \bibfield  {author} {\bibinfo {author} {\bibfnamefont {S.}~\bibnamefont {Toth}}\ and\ \bibinfo {author} {\bibfnamefont {B.}~\bibnamefont {Lake}},\ }\bibfield  {title} {\bibinfo {title} {Linear spin wave theory for single-q incommensurate magnetic structures},\ }\href {https://doi.org/10.1088/0953-8984/27/16/166002} {\bibfield  {journal} {\bibinfo  {journal} {Journal of Physics: Condensed Matter}\ }\textbf {\bibinfo {volume} {27}},\ \bibinfo {pages} {166002} (\bibinfo {year} {2015})}\BibitemShut {NoStop}%
\bibitem [{\citenamefont {Holstein}\ and\ \citenamefont {Primakoff}(1940)}]{HP}%
  \BibitemOpen
  \bibfield  {author} {\bibinfo {author} {\bibfnamefont {T.}~\bibnamefont {Holstein}}\ and\ \bibinfo {author} {\bibfnamefont {H.}~\bibnamefont {Primakoff}},\ }\bibfield  {title} {\bibinfo {title} {Field dependence of the intrinsic domain magnetization of a ferromagnet},\ }\href {https://doi.org/10.1103/PhysRev.58.1098} {\bibfield  {journal} {\bibinfo  {journal} {Phys. Rev.}\ }\textbf {\bibinfo {volume} {58}},\ \bibinfo {pages} {1098} (\bibinfo {year} {1940})}\BibitemShut {NoStop}%
\bibitem [{sup()}]{supp}%
  \BibitemOpen
  \href@noop {} {}\bibinfo {note} {See Supplemental Material at for a detailed derivation of magnon-phonon coupling Hamiltonian, exchange constants of monolayer $\mathrm{CrI_3}$, dependence of coupling constants on exchange-correlation functional, convergence test for TDDFT calculations and phonon properties.}\BibitemShut {Stop}%
\bibitem [{\citenamefont {Baroni}\ \emph {et~al.}(2001)\citenamefont {Baroni}, \citenamefont {de~Gironcoli}, \citenamefont {Dal~Corso},\ and\ \citenamefont {Giannozzi}}]{DFPT}%
  \BibitemOpen
  \bibfield  {author} {\bibinfo {author} {\bibfnamefont {S.}~\bibnamefont {Baroni}}, \bibinfo {author} {\bibfnamefont {S.}~\bibnamefont {de~Gironcoli}}, \bibinfo {author} {\bibfnamefont {A.}~\bibnamefont {Dal~Corso}},\ and\ \bibinfo {author} {\bibfnamefont {P.}~\bibnamefont {Giannozzi}},\ }\bibfield  {title} {\bibinfo {title} {Phonons and related crystal properties from density-functional perturbation theory},\ }\href {https://doi.org/10.1103/RevModPhys.73.515} {\bibfield  {journal} {\bibinfo  {journal} {Rev. Mod. Phys.}\ }\textbf {\bibinfo {volume} {73}},\ \bibinfo {pages} {515} (\bibinfo {year} {2001})}\BibitemShut {NoStop}%
\bibitem [{\citenamefont {Togo}\ \emph {et~al.}(2023)\citenamefont {Togo}, \citenamefont {Chaput}, \citenamefont {Tadano},\ and\ \citenamefont {Tanaka}}]{phonopy-phono3py-JPCM}%
  \BibitemOpen
  \bibfield  {author} {\bibinfo {author} {\bibfnamefont {A.}~\bibnamefont {Togo}}, \bibinfo {author} {\bibfnamefont {L.}~\bibnamefont {Chaput}}, \bibinfo {author} {\bibfnamefont {T.}~\bibnamefont {Tadano}},\ and\ \bibinfo {author} {\bibfnamefont {I.}~\bibnamefont {Tanaka}},\ }\bibfield  {title} {\bibinfo {title} {Implementation strategies in phonopy and phono3py},\ }\href {https://doi.org/10.1088/1361-648X/acd831} {\bibfield  {journal} {\bibinfo  {journal} {J. Phys. Condens. Matter}\ }\textbf {\bibinfo {volume} {35}},\ \bibinfo {pages} {353001} (\bibinfo {year} {2023})}\BibitemShut {NoStop}%
\bibitem [{\citenamefont {Chen}\ \emph {et~al.}(2018)\citenamefont {Chen}, \citenamefont {Chung}, \citenamefont {Gao}, \citenamefont {Chen}, \citenamefont {Stone}, \citenamefont {Kolesnikov}, \citenamefont {Huang},\ and\ \citenamefont {Dai}}]{Chen2018}%
  \BibitemOpen
  \bibfield  {author} {\bibinfo {author} {\bibfnamefont {L.}~\bibnamefont {Chen}}, \bibinfo {author} {\bibfnamefont {J.-H.}\ \bibnamefont {Chung}}, \bibinfo {author} {\bibfnamefont {B.}~\bibnamefont {Gao}}, \bibinfo {author} {\bibfnamefont {T.}~\bibnamefont {Chen}}, \bibinfo {author} {\bibfnamefont {M.~B.}\ \bibnamefont {Stone}}, \bibinfo {author} {\bibfnamefont {A.~I.}\ \bibnamefont {Kolesnikov}}, \bibinfo {author} {\bibfnamefont {Q.}~\bibnamefont {Huang}},\ and\ \bibinfo {author} {\bibfnamefont {P.}~\bibnamefont {Dai}},\ }\bibfield  {title} {\bibinfo {title} {{Topological Spin Excitations in Honeycomb Ferromagnet ${\mathrm{CrI}}_{3}$}},\ }\href {https://doi.org/10.1103/PhysRevX.8.041028} {\bibfield  {journal} {\bibinfo  {journal} {Phys. Rev. X}\ }\textbf {\bibinfo {volume} {8}},\ \bibinfo {pages} {041028} (\bibinfo {year} {2018})}\BibitemShut {NoStop}%
\bibitem [{\citenamefont {Gorni}\ \emph {et~al.}(2022)\citenamefont {Gorni}, \citenamefont {Baseggio}, \citenamefont {Delugas}, \citenamefont {Baroni},\ and\ \citenamefont {Timrov}}]{turboMagnon}%
  \BibitemOpen
  \bibfield  {author} {\bibinfo {author} {\bibfnamefont {T.}~\bibnamefont {Gorni}}, \bibinfo {author} {\bibfnamefont {O.}~\bibnamefont {Baseggio}}, \bibinfo {author} {\bibfnamefont {P.}~\bibnamefont {Delugas}}, \bibinfo {author} {\bibfnamefont {S.}~\bibnamefont {Baroni}},\ and\ \bibinfo {author} {\bibfnamefont {I.}~\bibnamefont {Timrov}},\ }\bibfield  {title} {\bibinfo {title} {turbomagnon – a code for the simulation of spin-wave spectra using the liouville-lanczos approach to time-dependent density-functional perturbation theory},\ }\href {https://doi.org/https://doi.org/10.1016/j.cpc.2022.108500} {\bibfield  {journal} {\bibinfo  {journal} {Computer Physics Communications}\ }\textbf {\bibinfo {volume} {280}},\ \bibinfo {pages} {108500} (\bibinfo {year} {2022})}\BibitemShut {NoStop}%
\bibitem [{\citenamefont {Ceperley}\ and\ \citenamefont {Alder}(1980)}]{LDA-CA}%
  \BibitemOpen
  \bibfield  {author} {\bibinfo {author} {\bibfnamefont {D.~M.}\ \bibnamefont {Ceperley}}\ and\ \bibinfo {author} {\bibfnamefont {B.~J.}\ \bibnamefont {Alder}},\ }\bibfield  {title} {\bibinfo {title} {Ground state of the electron gas by a stochastic method},\ }\href {https://doi.org/10.1103/PhysRevLett.45.566} {\bibfield  {journal} {\bibinfo  {journal} {Phys. Rev. Lett.}\ }\textbf {\bibinfo {volume} {45}},\ \bibinfo {pages} {566} (\bibinfo {year} {1980})}\BibitemShut {NoStop}%
\bibitem [{\citenamefont {Perdew}\ \emph {et~al.}(1996)\citenamefont {Perdew}, \citenamefont {Burke},\ and\ \citenamefont {Ernzerhof}}]{PBE}%
  \BibitemOpen
  \bibfield  {author} {\bibinfo {author} {\bibfnamefont {J.~P.}\ \bibnamefont {Perdew}}, \bibinfo {author} {\bibfnamefont {K.}~\bibnamefont {Burke}},\ and\ \bibinfo {author} {\bibfnamefont {M.}~\bibnamefont {Ernzerhof}},\ }\bibfield  {title} {\bibinfo {title} {Generalized gradient approximation made simple},\ }\href {https://doi.org/10.1103/PhysRevLett.77.3865} {\bibfield  {journal} {\bibinfo  {journal} {Phys. Rev. Lett.}\ }\textbf {\bibinfo {volume} {77}},\ \bibinfo {pages} {3865} (\bibinfo {year} {1996})}\BibitemShut {NoStop}%
\bibitem [{\citenamefont {Heyd}\ \emph {et~al.}(2003)\citenamefont {Heyd}, \citenamefont {Scuseria},\ and\ \citenamefont {Ernzerhof}}]{HSE}%
  \BibitemOpen
  \bibfield  {author} {\bibinfo {author} {\bibfnamefont {J.}~\bibnamefont {Heyd}}, \bibinfo {author} {\bibfnamefont {G.~E.}\ \bibnamefont {Scuseria}},\ and\ \bibinfo {author} {\bibfnamefont {M.}~\bibnamefont {Ernzerhof}},\ }\bibfield  {title} {\bibinfo {title} {Hybrid functionals based on a screened coulomb potential},\ }\href {https://doi.org/10.1063/1.1564060} {\bibfield  {journal} {\bibinfo  {journal} {The Journal of Chemical Physics}\ }\textbf {\bibinfo {volume} {118}},\ \bibinfo {pages} {8207} (\bibinfo {year} {2003})}\BibitemShut {NoStop}%
\bibitem [{\citenamefont {Ke}\ and\ \citenamefont {Katsnelson}(2021)}]{ke_electron_2021}%
  \BibitemOpen
  \bibfield  {author} {\bibinfo {author} {\bibfnamefont {L.}~\bibnamefont {Ke}}\ and\ \bibinfo {author} {\bibfnamefont {M.~I.}\ \bibnamefont {Katsnelson}},\ }\bibfield  {title} {\bibinfo {title} {Electron correlation effects on exchange interactions and spin excitations in {2D} van der {Waals} materials},\ }\href {https://doi.org/10.1038/s41524-020-00469-2} {\bibfield  {journal} {\bibinfo  {journal} {npj Computational Materials}\ }\textbf {\bibinfo {volume} {7}},\ \bibinfo {pages} {1} (\bibinfo {year} {2021})}\BibitemShut {NoStop}%
\bibitem [{\citenamefont {Lado}\ and\ \citenamefont {Fernández-Rossier}(2017)}]{Lado_2017}%
  \BibitemOpen
  \bibfield  {author} {\bibinfo {author} {\bibfnamefont {J.~L.}\ \bibnamefont {Lado}}\ and\ \bibinfo {author} {\bibfnamefont {J.}~\bibnamefont {Fernández-Rossier}},\ }\bibfield  {title} {\bibinfo {title} {On the origin of magnetic anisotropy in two dimensional cri3},\ }\href {https://doi.org/10.1088/2053-1583/aa75ed} {\bibfield  {journal} {\bibinfo  {journal} {2D Materials}\ }\textbf {\bibinfo {volume} {4}},\ \bibinfo {pages} {035002} (\bibinfo {year} {2017})}\BibitemShut {NoStop}%
\bibitem [{\citenamefont {Bl\"ochl}(1994)}]{Blochl1994}%
  \BibitemOpen
  \bibfield  {author} {\bibinfo {author} {\bibfnamefont {P.~E.}\ \bibnamefont {Bl\"ochl}},\ }\bibfield  {title} {\bibinfo {title} {Projector augmented-wave method},\ }\href {https://doi.org/10.1103/PhysRevB.50.17953} {\bibfield  {journal} {\bibinfo  {journal} {Phys. Rev. B}\ }\textbf {\bibinfo {volume} {50}},\ \bibinfo {pages} {17953} (\bibinfo {year} {1994})}\BibitemShut {NoStop}%
\bibitem [{\citenamefont {Kresse}\ and\ \citenamefont {Joubert}(1999)}]{VASP1}%
  \BibitemOpen
  \bibfield  {author} {\bibinfo {author} {\bibfnamefont {G.}~\bibnamefont {Kresse}}\ and\ \bibinfo {author} {\bibfnamefont {D.}~\bibnamefont {Joubert}},\ }\bibfield  {title} {\bibinfo {title} {From ultrasoft pseudopotentials to the projector augmented-wave method},\ }\href {https://doi.org/10.1103/PhysRevB.59.1758} {\bibfield  {journal} {\bibinfo  {journal} {Phys. Rev. B}\ }\textbf {\bibinfo {volume} {59}},\ \bibinfo {pages} {1758} (\bibinfo {year} {1999})}\BibitemShut {NoStop}%
\bibitem [{\citenamefont {Kresse}\ and\ \citenamefont {Furthm\"uller}(1996)}]{VASP2}%
  \BibitemOpen
  \bibfield  {author} {\bibinfo {author} {\bibfnamefont {G.}~\bibnamefont {Kresse}}\ and\ \bibinfo {author} {\bibfnamefont {J.}~\bibnamefont {Furthm\"uller}},\ }\bibfield  {title} {\bibinfo {title} {Efficient iterative schemes for ab initio total-energy calculations using a plane-wave basis set},\ }\href {https://doi.org/10.1103/PhysRevB.54.11169} {\bibfield  {journal} {\bibinfo  {journal} {Phys. Rev. B}\ }\textbf {\bibinfo {volume} {54}},\ \bibinfo {pages} {11169} (\bibinfo {year} {1996})}\BibitemShut {NoStop}%
\bibitem [{\citenamefont {Togo}(2023)}]{phonopy-phono3py-JPSJ}%
  \BibitemOpen
  \bibfield  {author} {\bibinfo {author} {\bibfnamefont {A.}~\bibnamefont {Togo}},\ }\bibfield  {title} {\bibinfo {title} {First-principles phonon calculations with phonopy and phono3py},\ }\href {https://doi.org/10.7566/JPSJ.92.012001} {\bibfield  {journal} {\bibinfo  {journal} {J. Phys. Soc. Jpn.}\ }\textbf {\bibinfo {volume} {92}},\ \bibinfo {pages} {012001} (\bibinfo {year} {2023})}\BibitemShut {NoStop}%
\bibitem [{\citenamefont {Dudarev}\ \emph {et~al.}(1998)\citenamefont {Dudarev}, \citenamefont {Botton}, \citenamefont {Savrasov}, \citenamefont {Humphreys},\ and\ \citenamefont {Sutton}}]{DFT+U}%
  \BibitemOpen
  \bibfield  {author} {\bibinfo {author} {\bibfnamefont {S.~L.}\ \bibnamefont {Dudarev}}, \bibinfo {author} {\bibfnamefont {G.~A.}\ \bibnamefont {Botton}}, \bibinfo {author} {\bibfnamefont {S.~Y.}\ \bibnamefont {Savrasov}}, \bibinfo {author} {\bibfnamefont {C.~J.}\ \bibnamefont {Humphreys}},\ and\ \bibinfo {author} {\bibfnamefont {A.~P.}\ \bibnamefont {Sutton}},\ }\bibfield  {title} {\bibinfo {title} {{Electron-energy-loss spectra and the structural stability of nickel oxide: An LSDA+U study}},\ }\href {https://doi.org/10.1103/PhysRevB.57.1505} {\bibfield  {journal} {\bibinfo  {journal} {Phys. Rev. B}\ }\textbf {\bibinfo {volume} {57}},\ \bibinfo {pages} {1505} (\bibinfo {year} {1998})}\BibitemShut {NoStop}%
\bibitem [{\citenamefont {Xiang}\ \emph {et~al.}(2011)\citenamefont {Xiang}, \citenamefont {Kan}, \citenamefont {Wei}, \citenamefont {Whangbo},\ and\ \citenamefont {Gong}}]{Xiang2011}%
  \BibitemOpen
  \bibfield  {author} {\bibinfo {author} {\bibfnamefont {H.~J.}\ \bibnamefont {Xiang}}, \bibinfo {author} {\bibfnamefont {E.~J.}\ \bibnamefont {Kan}}, \bibinfo {author} {\bibfnamefont {S.-H.}\ \bibnamefont {Wei}}, \bibinfo {author} {\bibfnamefont {M.-H.}\ \bibnamefont {Whangbo}},\ and\ \bibinfo {author} {\bibfnamefont {X.~G.}\ \bibnamefont {Gong}},\ }\bibfield  {title} {\bibinfo {title} {Predicting the spin-lattice order of frustrated systems from first principles},\ }\href {https://doi.org/10.1103/PhysRevB.84.224429} {\bibfield  {journal} {\bibinfo  {journal} {Phys. Rev. B}\ }\textbf {\bibinfo {volume} {84}},\ \bibinfo {pages} {224429} (\bibinfo {year} {2011})}\BibitemShut {NoStop}%
\bibitem [{\citenamefont {Giannozzi}\ \emph {et~al.}(2009)\citenamefont {Giannozzi}, \citenamefont {Baroni}, \citenamefont {Bonini}, \citenamefont {Calandra}, \citenamefont {Car}, \citenamefont {Cavazzoni}, \citenamefont {Ceresoli}, \citenamefont {Chiarotti}, \citenamefont {Cococcioni}, \citenamefont {Dabo}, \citenamefont {Corso}, \citenamefont {de~Gironcoli}, \citenamefont {Fabris}, \citenamefont {Fratesi}, \citenamefont {Gebauer}, \citenamefont {Gerstmann}, \citenamefont {Gougoussis}, \citenamefont {Kokalj}, \citenamefont {Lazzeri}, \citenamefont {Martin-Samos}, \citenamefont {Marzari}, \citenamefont {Mauri}, \citenamefont {Mazzarello}, \citenamefont {Paolini}, \citenamefont {Pasquarello}, \citenamefont {Paulatto}, \citenamefont {Sbraccia}, \citenamefont {Scandolo}, \citenamefont {Sclauzero}, \citenamefont {Seitsonen}, \citenamefont {Smogunov}, \citenamefont {Umari},\ and\ \citenamefont {Wentzcovitch}}]{QE1}%
  \BibitemOpen
  \bibfield  {author} {\bibinfo {author} {\bibfnamefont {P.}~\bibnamefont {Giannozzi}}, \bibinfo {author} {\bibfnamefont {S.}~\bibnamefont {Baroni}}, \bibinfo {author} {\bibfnamefont {N.}~\bibnamefont {Bonini}}, \bibinfo {author} {\bibfnamefont {M.}~\bibnamefont {Calandra}}, \bibinfo {author} {\bibfnamefont {R.}~\bibnamefont {Car}}, \bibinfo {author} {\bibfnamefont {C.}~\bibnamefont {Cavazzoni}}, \bibinfo {author} {\bibfnamefont {D.}~\bibnamefont {Ceresoli}}, \bibinfo {author} {\bibfnamefont {G.~L.}\ \bibnamefont {Chiarotti}}, \bibinfo {author} {\bibfnamefont {M.}~\bibnamefont {Cococcioni}}, \bibinfo {author} {\bibfnamefont {I.}~\bibnamefont {Dabo}}, \bibinfo {author} {\bibfnamefont {A.~D.}\ \bibnamefont {Corso}}, \bibinfo {author} {\bibfnamefont {S.}~\bibnamefont {de~Gironcoli}}, \bibinfo {author} {\bibfnamefont {S.}~\bibnamefont {Fabris}}, \bibinfo {author} {\bibfnamefont {G.}~\bibnamefont {Fratesi}}, \bibinfo {author} {\bibfnamefont {R.}~\bibnamefont {Gebauer}}, \bibinfo {author} {\bibfnamefont
  {U.}~\bibnamefont {Gerstmann}}, \bibinfo {author} {\bibfnamefont {C.}~\bibnamefont {Gougoussis}}, \bibinfo {author} {\bibfnamefont {A.}~\bibnamefont {Kokalj}}, \bibinfo {author} {\bibfnamefont {M.}~\bibnamefont {Lazzeri}}, \bibinfo {author} {\bibfnamefont {L.}~\bibnamefont {Martin-Samos}}, \bibinfo {author} {\bibfnamefont {N.}~\bibnamefont {Marzari}}, \bibinfo {author} {\bibfnamefont {F.}~\bibnamefont {Mauri}}, \bibinfo {author} {\bibfnamefont {R.}~\bibnamefont {Mazzarello}}, \bibinfo {author} {\bibfnamefont {S.}~\bibnamefont {Paolini}}, \bibinfo {author} {\bibfnamefont {A.}~\bibnamefont {Pasquarello}}, \bibinfo {author} {\bibfnamefont {L.}~\bibnamefont {Paulatto}}, \bibinfo {author} {\bibfnamefont {C.}~\bibnamefont {Sbraccia}}, \bibinfo {author} {\bibfnamefont {S.}~\bibnamefont {Scandolo}}, \bibinfo {author} {\bibfnamefont {G.}~\bibnamefont {Sclauzero}}, \bibinfo {author} {\bibfnamefont {A.~P.}\ \bibnamefont {Seitsonen}}, \bibinfo {author} {\bibfnamefont {A.}~\bibnamefont {Smogunov}}, \bibinfo {author}
  {\bibfnamefont {P.}~\bibnamefont {Umari}},\ and\ \bibinfo {author} {\bibfnamefont {R.~M.}\ \bibnamefont {Wentzcovitch}},\ }\bibfield  {title} {\bibinfo {title} {Quantum espresso: a modular and open-source software project for quantum simulations of materials},\ }\href {https://doi.org/10.1088/0953-8984/21/39/395502} {\bibfield  {journal} {\bibinfo  {journal} {Journal of Physics: Condensed Matter}\ }\textbf {\bibinfo {volume} {21}},\ \bibinfo {pages} {395502} (\bibinfo {year} {2009})}\BibitemShut {NoStop}%
\bibitem [{\citenamefont {Giannozzi}\ \emph {et~al.}(2017)\citenamefont {Giannozzi}, \citenamefont {Andreussi}, \citenamefont {Brumme}, \citenamefont {Bunau}, \citenamefont {Nardelli}, \citenamefont {Calandra}, \citenamefont {Car}, \citenamefont {Cavazzoni}, \citenamefont {Ceresoli}, \citenamefont {Cococcioni}, \citenamefont {Colonna}, \citenamefont {Carnimeo}, \citenamefont {Corso}, \citenamefont {de~Gironcoli}, \citenamefont {Delugas}, \citenamefont {DiStasio}, \citenamefont {Ferretti}, \citenamefont {Floris}, \citenamefont {Fratesi}, \citenamefont {Fugallo}, \citenamefont {Gebauer}, \citenamefont {Gerstmann}, \citenamefont {Giustino}, \citenamefont {Gorni}, \citenamefont {Jia}, \citenamefont {Kawamura}, \citenamefont {Ko}, \citenamefont {Kokalj}, \citenamefont {Küçükbenli}, \citenamefont {Lazzeri}, \citenamefont {Marsili}, \citenamefont {Marzari}, \citenamefont {Mauri}, \citenamefont {Nguyen}, \citenamefont {Nguyen}, \citenamefont {de-la Roza}, \citenamefont {Paulatto}, \citenamefont {Poncé},
  \citenamefont {Rocca}, \citenamefont {Sabatini}, \citenamefont {Santra}, \citenamefont {Schlipf}, \citenamefont {Seitsonen}, \citenamefont {Smogunov}, \citenamefont {Timrov}, \citenamefont {Thonhauser}, \citenamefont {Umari}, \citenamefont {Vast}, \citenamefont {Wu},\ and\ \citenamefont {Baroni}}]{QE2}%
  \BibitemOpen
  \bibfield  {author} {\bibinfo {author} {\bibfnamefont {P.}~\bibnamefont {Giannozzi}}, \bibinfo {author} {\bibfnamefont {O.}~\bibnamefont {Andreussi}}, \bibinfo {author} {\bibfnamefont {T.}~\bibnamefont {Brumme}}, \bibinfo {author} {\bibfnamefont {O.}~\bibnamefont {Bunau}}, \bibinfo {author} {\bibfnamefont {M.~B.}\ \bibnamefont {Nardelli}}, \bibinfo {author} {\bibfnamefont {M.}~\bibnamefont {Calandra}}, \bibinfo {author} {\bibfnamefont {R.}~\bibnamefont {Car}}, \bibinfo {author} {\bibfnamefont {C.}~\bibnamefont {Cavazzoni}}, \bibinfo {author} {\bibfnamefont {D.}~\bibnamefont {Ceresoli}}, \bibinfo {author} {\bibfnamefont {M.}~\bibnamefont {Cococcioni}}, \bibinfo {author} {\bibfnamefont {N.}~\bibnamefont {Colonna}}, \bibinfo {author} {\bibfnamefont {I.}~\bibnamefont {Carnimeo}}, \bibinfo {author} {\bibfnamefont {A.~D.}\ \bibnamefont {Corso}}, \bibinfo {author} {\bibfnamefont {S.}~\bibnamefont {de~Gironcoli}}, \bibinfo {author} {\bibfnamefont {P.}~\bibnamefont {Delugas}}, \bibinfo {author} {\bibfnamefont
  {R.~A.}\ \bibnamefont {DiStasio}}, \bibinfo {author} {\bibfnamefont {A.}~\bibnamefont {Ferretti}}, \bibinfo {author} {\bibfnamefont {A.}~\bibnamefont {Floris}}, \bibinfo {author} {\bibfnamefont {G.}~\bibnamefont {Fratesi}}, \bibinfo {author} {\bibfnamefont {G.}~\bibnamefont {Fugallo}}, \bibinfo {author} {\bibfnamefont {R.}~\bibnamefont {Gebauer}}, \bibinfo {author} {\bibfnamefont {U.}~\bibnamefont {Gerstmann}}, \bibinfo {author} {\bibfnamefont {F.}~\bibnamefont {Giustino}}, \bibinfo {author} {\bibfnamefont {T.}~\bibnamefont {Gorni}}, \bibinfo {author} {\bibfnamefont {J.}~\bibnamefont {Jia}}, \bibinfo {author} {\bibfnamefont {M.}~\bibnamefont {Kawamura}}, \bibinfo {author} {\bibfnamefont {H.-Y.}\ \bibnamefont {Ko}}, \bibinfo {author} {\bibfnamefont {A.}~\bibnamefont {Kokalj}}, \bibinfo {author} {\bibfnamefont {E.}~\bibnamefont {Küçükbenli}}, \bibinfo {author} {\bibfnamefont {M.}~\bibnamefont {Lazzeri}}, \bibinfo {author} {\bibfnamefont {M.}~\bibnamefont {Marsili}}, \bibinfo {author} {\bibfnamefont
  {N.}~\bibnamefont {Marzari}}, \bibinfo {author} {\bibfnamefont {F.}~\bibnamefont {Mauri}}, \bibinfo {author} {\bibfnamefont {N.~L.}\ \bibnamefont {Nguyen}}, \bibinfo {author} {\bibfnamefont {H.-V.}\ \bibnamefont {Nguyen}}, \bibinfo {author} {\bibfnamefont {A.~O.}\ \bibnamefont {de-la Roza}}, \bibinfo {author} {\bibfnamefont {L.}~\bibnamefont {Paulatto}}, \bibinfo {author} {\bibfnamefont {S.}~\bibnamefont {Poncé}}, \bibinfo {author} {\bibfnamefont {D.}~\bibnamefont {Rocca}}, \bibinfo {author} {\bibfnamefont {R.}~\bibnamefont {Sabatini}}, \bibinfo {author} {\bibfnamefont {B.}~\bibnamefont {Santra}}, \bibinfo {author} {\bibfnamefont {M.}~\bibnamefont {Schlipf}}, \bibinfo {author} {\bibfnamefont {A.~P.}\ \bibnamefont {Seitsonen}}, \bibinfo {author} {\bibfnamefont {A.}~\bibnamefont {Smogunov}}, \bibinfo {author} {\bibfnamefont {I.}~\bibnamefont {Timrov}}, \bibinfo {author} {\bibfnamefont {T.}~\bibnamefont {Thonhauser}}, \bibinfo {author} {\bibfnamefont {P.}~\bibnamefont {Umari}}, \bibinfo {author}
  {\bibfnamefont {N.}~\bibnamefont {Vast}}, \bibinfo {author} {\bibfnamefont {X.}~\bibnamefont {Wu}},\ and\ \bibinfo {author} {\bibfnamefont {S.}~\bibnamefont {Baroni}},\ }\bibfield  {title} {\bibinfo {title} {Advanced capabilities for materials modelling with quantum espresso},\ }\href {https://doi.org/10.1088/1361-648X/aa8f79} {\bibfield  {journal} {\bibinfo  {journal} {Journal of Physics: Condensed Matter}\ }\textbf {\bibinfo {volume} {29}},\ \bibinfo {pages} {465901} (\bibinfo {year} {2017})}\BibitemShut {NoStop}%
\bibitem [{\citenamefont {Hamann}(2013)}]{ONCV1}%
  \BibitemOpen
  \bibfield  {author} {\bibinfo {author} {\bibfnamefont {D.~R.}\ \bibnamefont {Hamann}},\ }\bibfield  {title} {\bibinfo {title} {Optimized norm-conserving vanderbilt pseudopotentials},\ }\href {https://doi.org/10.1103/PhysRevB.88.085117} {\bibfield  {journal} {\bibinfo  {journal} {Phys. Rev. B}\ }\textbf {\bibinfo {volume} {88}},\ \bibinfo {pages} {085117} (\bibinfo {year} {2013})}\BibitemShut {NoStop}%
\bibitem [{\citenamefont {{van Setten}}\ \emph {et~al.}(2018)\citenamefont {{van Setten}}, \citenamefont {Giantomassi}, \citenamefont {Bousquet}, \citenamefont {Verstraete}, \citenamefont {Hamann}, \citenamefont {Gonze},\ and\ \citenamefont {Rignanese}}]{van2018pseudodojo}%
  \BibitemOpen
  \bibfield  {author} {\bibinfo {author} {\bibfnamefont {M.}~\bibnamefont {{van Setten}}}, \bibinfo {author} {\bibfnamefont {M.}~\bibnamefont {Giantomassi}}, \bibinfo {author} {\bibfnamefont {E.}~\bibnamefont {Bousquet}}, \bibinfo {author} {\bibfnamefont {M.}~\bibnamefont {Verstraete}}, \bibinfo {author} {\bibfnamefont {D.}~\bibnamefont {Hamann}}, \bibinfo {author} {\bibfnamefont {X.}~\bibnamefont {Gonze}},\ and\ \bibinfo {author} {\bibfnamefont {G.-M.}\ \bibnamefont {Rignanese}},\ }\bibfield  {title} {\bibinfo {title} {The pseudodojo: Training and grading a 85 element optimized norm-conserving pseudopotential table},\ }\href {https://doi.org/https://doi.org/10.1016/j.cpc.2018.01.012} {\bibfield  {journal} {\bibinfo  {journal} {Computer Physics Communications}\ }\textbf {\bibinfo {volume} {226}},\ \bibinfo {pages} {39} (\bibinfo {year} {2018})}\BibitemShut {NoStop}%
\end{thebibliography}

\end{document}